\documentclass{article}

\usepackage{arxiv}

\usepackage[utf8]{inputenc} 
\usepackage[T1]{fontenc}    
\usepackage{hyperref}       
\usepackage{url}            
\usepackage{booktabs}       
\usepackage{amsfonts}       
\usepackage{nicefrac}       
\usepackage{microtype}      
\usepackage{graphicx}
\usepackage{doi}
\usepackage{amssymb}
\usepackage{latexsym}
\usepackage{epsfig}
\usepackage{amsmath}

\def\convinlaw{\stackrel{{\cal L}}{\Longrightarrow }}

\def\tends{\rightarrow}

\newcommand{\R}{{\rm I}\!{\rm R}}

\newtheorem{theorem}{Theorem}[section]

\newtheorem{proposition}{Proposition}[section]
\newtheorem{corollary}{Corollary}[section]
\newtheorem{Remark}{Remark}[section]

\newtheorem{algorithm}{Algorithm}[section]

\title{Estimating the Spectral Density at Frequencies Near Zero }

\author{ Tucker S. McElroy\\U.S. Census Bureau \\
  4600 Silver Hill Road, Washington, DC 20233\\
\texttt{tucker.s.mcelroy@census.gov} \And 
       Dimitris N. Politis \\Department of
Mathematics and Halicioglu Data Science Institute  \\
University of California at San Diego \\ La Jolla, CA 92093-0112, USA
      \texttt{dpolitis@ucsd.edu} }

\date{}

\hypersetup{
pdftitle={Estimating the Spectral Density at Frequencies Near Zero},
pdfsubject={},
pdfauthor={Tucker S.~McElroy, Dimitris N.~Politis},
pdfkeywords={Flat-top lag-windows, Function Estimation,  Kernel smoothing,
  Local Polynomials, Long-run variance, Sample mean},
}

\begin{document}
\maketitle

 \begin{abstract}  Estimating the spectral density function
$f(w)$ for some  $w\in [-\pi, \pi]$ has been traditionally performed
by kernel smoothing the periodogram and related techniques.
Kernel smoothing is tantamount to local averaging, i.e., approximating
$f(w)$ by a constant over a window of small width.
Although $f(w)$ is uniformly continuous and periodic with period $2\pi$,
 in this paper we recognize the fact that  $w=0$   effectively
acts as a boundary point in the underlying kernel smoothing problem, and
the same is true for   $w=\pm \pi$.
It is well-known that local averaging may be  suboptimal in kernel regression
at (or near) a boundary point. As an alternative, we propose a
local polynomial regression of the periodogram or log-periodogram
when $w$ is at (or near) the points 0 or $\pm \pi$.
The case $w=0$ is of particular importance  since $f(0)$ is the large-sample
variance of the sample mean; hence,  estimating $f(0)$ is crucial in order to conduct
any sort of inference on the mean.
\end{abstract}



\section{Introduction}

Many applications of
time series analysis involve the nonparametric estimation
  of the spectral density function $f(w)$ as some point $w\in [-\pi, \pi]$; examples include
astronomy, economics, electrical engineering, physics, etc.
The prevalent spectral estimation  method in the literature
goes back to Bartlett (1946, 1948) and Daniell (1946),
and can be represented in two equivalent ways:
(a) kernel smoothing of the periodogram,  and
(b)  tapered Fourier series of the sample autocovariances.

 By the late 1950s the subject was
already well understood, as the early books
by Grenander  and   Rosenblatt  (1957), and
  Blackman  and Tukey (1959) demonstrate.
Influential papers at the time include Hannan  (1957, 1958),
 Parzen (1957, 1961), and Priestley  (1962).
More recent advances include smoothing the log-periodogram of
Wahba (1980), smoothing with the trapezoidal `flat-top' lag-window
of Politis and Romano (1995),  local polynomial smoothing of  the periodogram by
 Fan  and Kreutzberger  (1998), and the  thresholded Fourier series of
Paparoditis  and    Politis  (2012).
The book-length treatments in
Hannan (1970), Brillinger (1981),  Priestley (1981),
Brockwell and Davis (1991),   Rosenblatt (1985),
  Percival  and Walden (1993), and McElroy and Politis (2020)
contain a vast number of additional references.

Kernel smoothing of the periodogram is tantamount to local averaging, i.e., approximating
$f(w)$ by a constant over a window of small width.
Although the spectral density function $f(w)$ is uniformly continuous and periodic with period $2\pi$,
 in this paper we recognize the fact that   $w=0$   effectively
acts as a boundary point in the underlying kernel smoothing problem, and
the same is true for   $w=\pm \pi$.
It is well-known that local averaging may be suboptimal in kernel regression
at (or near) a boundary point. As an alternative, we propose a
local polynomial regression of the periodogram (or log-periodogram)
when $w$ is at (or near) the points 0 or $\pm \pi$.
The case $w=0$ is of particular importance  since $f(0)$ is the large-sample
variance of the sample mean; hence,  estimating $f(0)$ is crucial in order to conduct
inference on the mean.
Due to the symmetries of $f(w)$, the recommended
polynomial regression at the two boundaries entails a quadratic without a linear term.

The remainder of the paper is structured as follows.
In Section \ref{sec.2.1},   traditional  spectral estimators are reviewed,
and the new method is proposed in  Section \ref{sec.2.2}.
Section \ref{sec.2.3} describes how the same ideas are applicable to
spectral   estimation near the two boundaries.
Section   \ref{sec.3} studies the asymptotic performance of the new proposal,
and calculates the optimal bandwidth.
  Section   \ref{sec.2.4} discusses the positivity issue and proposes a
log-periodogram estimator that is designed to positive, while
 Section   \ref{sec.2.5} shows how to combine the boundary estimators
with a traditional spectral estimator valid for interior points.
Section \ref{sec.4.3} makes extensions to nonstationary time series with
time-varying mean function. 
 Section   \ref{sec.sims} presents our proposed method of 
data-based bandwith choice, and includes results of an extensive
  finite-sample simulation experiment; the simulations required   
several months
of computer time  on a 20-core  workstation dedicated to statistical computing.
Finally,  Section   \ref{se:gross} gives two  real data applications: 
the U.S. Gross Domestic Product (GDP) and the Global Land-Ocean Temperature
Index (GLOTI).
An array of tables and figures is deferred to the Supplement (\url{https://mathweb.ucsd.edu/~politis/PAPER/LocalQuadSupp.pdf}).

\section{Spectral density estimation}
\label{sec.2}

Suppose $X_1,\ldots,X_n$ are observations from
the  (strictly)  stationary real-valued
sequence $\{X_t, t\in {\bf Z} \}$ having   mean $\mu = EX_t$ and autocovariance
sequence $\gamma (k) =E(X_t -\mu )(X_{t+|k|} -\mu )$,
   where both $\mu$ and $\gamma (\cdot )$ are unknown;
also define the autocorrelation sequence $  \rho (k)=   \gamma (k)/  \gamma (0)$.
Typical estimators of $\mu$ and $\gamma (k)$ are the sample mean
$\bar X_n=n^{-1}\sum_{i=1}^{n} X_i$ and sample autocovariance
$\hat \gamma (k)=n^{-1}\sum_{i=1}^{n-|k|} (X_i- \bar X_n)(X_{i+|k|}- \bar X_n)  $ respectively;
 $\hat \gamma (k)  $ is defined to be
zero when  $|k|\geq n$.
Our objective is the nonparametric estimation of
 the spectral density function
 $f(w)=   \sum_{s=-\infty}^\infty e^{iws}\gamma (s) $ 
assuming that the infinite sum is well-defined.
Consider the assumption:

\vskip .1in
\noindent
{\bf Assumption A$(p)$}: The spectral density 
 $f(w)$ is $p$-times continuously differentiable for all real $w$.
\vskip .1in
  From here on and throughout this paper, we will   assume {\it Assumption A($p$) holds for some  $p\geq 2$.}
 Note that    $f(w) $
 (and its derivatives) are periodic functions  with period $2\pi$,
and $f(w)$ has even symmetry,
i.e., $f(w)=f(-w)$; hence, we can focus on estimating it for $w\in  [0, \pi]$ only.
It is easy to see that if $\sum_{s=-\infty}^\infty |s|^p|\gamma (s)|<\infty  $ for some
non-negative  integer $p$,  then Assumption A$(p)$ holds true;
see e.g.  Proposition 6.1.5 of McElroy and Politis (2020).

\subsection{Traditional spectral   estimation }
 \label{sec.2.1}

 The traditional   kernel estimator of  $f(w)$  in its Fourier series form
is defined as
\begin{equation} \widehat{ f}  (w)=  \sum_{s=-\infty}^\infty e^{iws}\lambda  (s/M) \widehat{ \gamma } (s),
\label{eq.lwse}
\end{equation}
where  the {\it lag-window} $\lambda  (s) $ is  a bounded, square-integrable function with even symmetry
 that satisfies $\lambda  (0) =1$. 
Suppose that $\lambda  (s) $ has   $q$ continuous derivatives at the origin;
if $\lambda ^{(k)} (0)=0$ for $k=1,\ldots, q-1$, then $\lambda  (s)$ is said to have
  {\it order}  equal to $q$. 
 Popular  choices for 2nd order lag-windows
have been proposed by   Parzen (1961)
and Priestley    (1962).

The periodogram is defined as $I(w)=  \sum_{s=-\infty}^\infty e^{iws} \widehat{ \gamma} (s) $.
  By the convolution formula --- see
Proposition 6.1.11 of McElroy and Politis (2020) --- it follows that we can alternatively express
$\widehat{ f } (w)$ by means of smoothing the periodogram, i.e.,
\begin{equation} \widehat{ f}  (w)=
\frac{1}{2\pi}\int_{-\pi}^\pi   \Lambda_M(s) I(w+s)ds,
\label{eq.lwseK}
\end{equation}
where $\Lambda_M(w)= \sum_{s=-\infty}^\infty e^{iws}\lambda  (s/M)$
is the so-called spectral 
 {\it kernel}.
The integral in eq. (\ref{eq.lwseK}) is practically approximated by a Riemann
sum over the Fourier frequencies $w_j=2\pi j/n$, i.e.,
\begin{equation} \hat f  (w)\approx
\frac{1}{n} \sum_{j\in J_n}    \Lambda_M(w_j) I(w+w_j),
\label{eq.lwseKapprox}
\end{equation}
where the error in the approximation is $O(1/n)$.  The index set $J_n$ consists of $n$
consecutive integers as follows:
$J_n=\{-\frac{n-1}{2},\ldots, 0,\ldots,  \frac{n-1}{2} \}$ if $n$ is odd, and
$J_n=\{-\frac{n}{2}+1,\ldots,0,\ldots,  \frac{n}{2} \}$  if $n$ is even.

It is well known that under regularity conditions
\begin{equation}
Var [ \hat f  (w) ] = \eta (w) f^2(w) \frac{M}{n} \int_{-\infty}^\infty \lambda^2(s)ds + o \left( \frac{M}{n} \right)
\label{eq.fVar}
\end{equation}
as $n\to \infty$, $M\to \infty$ but $\frac{M}{n}\to 0$; see e.g. Rosenblatt  (1984),
 Shao and Wu (2007), or
 Ch. 9 of McElroy and Politis (2020).  The function $\eta (w)=2 $ if $w$ is zero or an integer
multiple of $\pi$; otherwise, $\eta (w)=1 $.  Furthermore, under Assumption A$(p)$ for some $p\geq 2$ we have
\begin{equation}
Bias[ \hat f  (w) ] = O\left( M^{-\min (p,q) }\right),
\label{eq.fBias}
\end{equation}
where $q$ is the order of the kernel. It is apparent that if
$f  (w)$ is very smooth, i.e., possessing a large number of derivatives, it would be advantageous
to use a high-order kernel.

A more recent development is the use
  of {\it flat-top} lag-windows; see
Paradigm 9.9.1 of McElroy and Politis (2020).
Flat-top lag-windows    have been
  shown to achieve the optimal rate of convergence in
a given smoothness class by automatically
{\it adapting} to the underlying
smoothness of the true spectral density.
A general `flat-top'  lag-window is defined as
\begin{equation}
\lambda_{g,c} (x) = \begin{cases}
    1 \quad \mbox{if} \, |x|\leq c  \\
      g(x)  \quad \mbox{else};  
\end{cases}
\label{eq.lambda-gc}
\end{equation}
here $c>0$ is a shape parameter,
and  $g: \R \to [-1,1]$ is a symmetric  function,
  continuous at all but a finite number of points, and
satisfying $g(c)=1$,   and $\int_c^\infty g^2(x)dx <
\infty$.
Note that a flat-top lag-window has infinite order, as all its derivatives vanish at the origin.
 The trapezoidal lag-window of Politis and Romano (1995), given by
  $\lambda  (x) = \left( \min \{1, 2(1-  |x| ) \} \right)^+$,
is a prominent member of the flat-top family;
here, $(x)^+=\max (x,0)$.

It has been found---see e.g. Politis (2001, 2003, 2011)---that
estimator (\ref{eq.lwse}) using a flat-top lag-window
achieves the optimal rate of convergence in a given smoothness
class;  these optimal rates are delineated in Samarov (1977).
Hence, a flat-top lag-window
estimator  exhibits {\it adaptivity} to the
(unknown) degree of smoothness of the underlying true spectral density;
the degree of smoothness can  be quantified by the rate of the decay of the
autocovariance.
Perhaps more importantly, these optimal rates are (almost) achieved even
when an empirical data-dependent choice of the bandwidth $M$
is employed---see~Politis (2003, 2011).

\begin{Remark} \rm
A minor complication  is that   flat-top lag-windows are not positive definite,
i.e., using them does not ensure that $
 \hat f  (w)$ will be non-negative  almost surely.
Since  $f(w)\geq 0$ always, an easy fix is to take the positive part of $\hat f  (w)$,
i.e., define the modified estimator $\hat f ^+ (w)= \max (\hat f  (w),0)$.
Alternative corrections for positivity are discussed in Appendix A of
McMurry and  Politis~(2015).
\label{Re.hat f plus}
\end{Remark}

\subsection{Improved spectral   estimation at the origin}
\label{sec.2.2}

Consider for a moment a different setup,  where we have
data $Y_1, \ldots, Y_n$ from the  nonparametric regression
model
\begin{equation}
 Y_j=f(x_j)+\sigma (x_j) \epsilon_j,
\label{eq.npreg}
\end{equation}
where the errors $\epsilon_i$
are independent, identically distributed (i.i.d.) with mean zero, and variance one.
Here,  $f(x)$ and $\sigma (x)$ are  smooth (but otherwise
unknown) functions with compact  support $[c_1, c_2]$.
Kernel smoothing, i.e., local averaging of the $Y_i$ data, is a fundamental
way to estimate the function $f(x)$ for some  $x\in [c_1, c_2]$.

Local averaging is tantamount to approximating $f(x)$ by a constant over a small
window of the type $[x-h, x+h]$; here, $h$ is a small positive number called the {\it bandwidth}.
It is well known that local linear (and local polynomial) fitting outperforms
local averaging when either the point of interest $x$ is in a region where few
design points $x_j$ are found, or $x$ is near (or on) one of  the boundary points $ c_1, c_2 $;
see Fan and Gijbels  (1996) and the references therein.

Recall that the periodogram ordinates $I(w_j)$ for $j \in  J_n$ are approximately independent.
The   asymptotic distribution of $I(w)/ f(w) $ is Exponential with mean one if $w\in (0,\pi)$;
if $w=\pm \pi$, then $I(w)/ f(w) $ is  asymptotically chi-square with one degree of freedom.
Note that $I(0)=0$ identically; see Corollary 9.7.5 of McElroy and Politis (2020).
That said, it is apparent that eq. (\ref{eq.lwseKapprox}) is tantamount
to applying local averaging on
the nonparametric regression
(\ref{eq.npreg}) with $Y_j= I(w_j)$,  $\sigma (w_j) =f (w_j)$, and
the $\epsilon_j$ being (approximately) i.i.d.~Exponential with mean one.

Although Fan  and Kreutzberger   (1998) did work out the local polynomial method of
estimating $f(w) $  for  $w\in (0,\pi)$, their results were not appreciably better than
the local averaging of eq. (\ref{eq.lwseKapprox}). The intuitive  reason is that the design points
are uniformly spaced $w_j=2\pi j/n$,  which makes local linear fitting equivalent to simple
local averaging.

Since $f(w) $ is   continuous over    $[-\infty,\infty]$ and
periodic with period $2\pi$,
 it has been thought
that its nonparametric estimation does not suffer from boundary effects; this, however,  is not entirely correct.
To see why, focus on the case $w=0$; since $I(0)=0$, the estimator
$\hat f(0)$ from eq. (\ref{eq.lwseKapprox}) involves averaging periodogram ordinates to the left and to the
right of the origin. But
 $I(w) $ has even symmetry, and is periodic with period $2\pi$ as well.
Because of the symmetry $I(w)  =I(-w) $, the estimator
$\hat f(0)$ from eq. (\ref{eq.lwseKapprox}) is tantamount to a {\it one-sided} average, i.e.,
\begin{equation} \hat f  (0)\approx
\frac{1}{n} \sum_{j\in J_n}    \Lambda_M(w_j) I( w_j)
=\frac{2}{n} \sum_{j\in J_n^+}    \Lambda_M(w_j) I( w_j),
\label{eq.one sided}
\end{equation}
where $J_n^+$ is the subset of positive indices belonging to $J_n$; this
actually explains the doubling of the variance when $w=0$ in eq. (\ref{eq.fVar})
since the effective sample size is halved due to the symmetry.

Therefore, $w=0$ effectively acts as a {\it  boundary} point in nonparametric regression of
periodogram ordinates. The same is true for the points $w=\pm \pi$ due to the periodicity
of $I(w) $. Hence, we may propose local polynomial fitting of periodogram 
ordinates in order to estimate $f(w) $ when $w$ is zero, or an integer multiple of $\pi$.
Because of the even symmetry (and periodicity) of $f(w) $, local  linear fitting would again
be equivalent to local averaging when $w=0 $ or $\pm \pi$. A local quadratic (without a linear term)
can be fitted instead;  setting  $h = 2  \pi \delta$, for a small  $\delta \in (0,.5)$, we may use the Taylor approximations
\begin{equation}
 f(w)\approx a_0+b_0w^2  \  \  \mbox{for} \  \ w\in [-  2  \pi \delta,  2  \pi \delta]
\label{eq.loc-quad 0}
\end{equation}
\begin{equation}
\mbox{and} \ \  f(w)\approx a_1+b_1(w-\pi) ^2   \  \  \mbox{for} \  \ w\in [\pi- 2  \pi \delta, \pi+ 2  \pi \delta].
\label{eq.loc-quad pi}
\end{equation}
 Since $f(-\pi) $=$f( \pi) $, we do not have to address
the case $w=-\pi$ separately.
\vskip .1in
Our proposal   then is summarized in the  following algorithm:
\begin{algorithm}
\label{algo.1}
\vskip .1in
\end{algorithm}

\begin{enumerate}
\item [(i)]
Let $\hat a_0$ and $\hat b_0$
be  the estimators of  $a_0$ and $b_0$  in eq. (\ref{eq.loc-quad 0}) based on  a
(possibly weighted)   regression of $I(w )$ on
  an intercept  $a_0$ and quadratic term $b_0w^2$. The data to be used in this regression
are $I(w_1), \ldots, I(w_m) $, where $w_m$ is the largest Fourier frequency less or equal to $2  \pi \delta$;
since $w_m=2\pi m/n \leq 2  \pi \delta$, it follows that $m=[\delta n ]$, where $[\cdot]$
denotes the integer part. Explicit 
formulas for $\hat{a}_0$ and $\hat{b}_0$ are provided in Section \ref{sec.3}.

\item [(ii)]
Similarly,  $\hat a_1$ and $\hat b_1$
are  the estimators of  $a_1$ and $b_1$  in eq. (\ref{eq.loc-quad pi}) based on a
(possibly weighted)   regression of $I(w )$ on
  an intercept  $a_1$ and quadratic term $b_1(w-\pi) ^2$. The data to be used in this regression
are the $I(w_j)  $  with indices corresponding to the $m$ largest elements of set $J_n$.

\item [(iii)] Finally,  $\tilde f(0)=\hat a_0$  is  the proposed new  estimator of $  f(0)$,
and  $\tilde f(\pi) =\hat a_1$ is  the proposed  estimator of $  f(\pi)$.
As will be shown in Section \ref{sec.3}, to achieve consistent estimation
in either of the above cases,
 we would need $\delta \to 0$ but $\delta n\to \infty $ as $n\to \infty $, which is equivalent to
$m\to \infty $ but with $m/n \to 0$.
\end{enumerate}
 
\begin{Remark} \rm  
Marron  and Ruppert  (1994) 
 studied in detail the problem of nonparametric   estimation of a probability density 
  at its boundaries.   The case of the nonparametric  spectral density estimator 
  $ \widehat{ f}  (w)$ using a non-negative spectral kernel   $\Lambda_M$ is completely analogous. 
  In both cases, the bias of $ \widehat{ f}  (w)$ is of order $f^{\prime \prime } (w)/M^2$ if 
  $w$ is an interior point. Furthermore, the first term in the expansion of 
  the bias of $ \widehat{ f}  (\theta)$ is of order 
      $f^{\prime   } (\theta)/M$,  where $\theta$ is one of the
  ``boundary''  points, i.e., $\theta =0$ or $\pm \pi$.  
  Note, however, that because of the even symmetry (and periodicity) of $f(w) $, we have
$f^\prime (\theta)=0$. Therefore, the bias of $ \widehat{ f}  (w)$ is of order $O(M^{-2})$ 
for all $w$, {\it even at the boundary}. It is perhaps for this reason that spectral   estimation has not been identified in the past literature as having a boundary issue. 
 Indeed, the situation at the
boundary could be better described as a boundary {\it opportunity} rather than a boundary problem.  
 
Recall that we motivated the presence of the boundary issue by the fact that
the variance of  $ \widehat{ f}  (w)$ is doubled when  $w$ is at the boundary;
however,  there is an issue with the bias as well. Note that 
$|Bias ( \widehat{ f}  (w)) \approx |f^{\prime \prime } (w)|/M^2$;
furthermore, $ |f^{\prime \prime } (w)|$ has a (local) maximum when
$w =0$ or $\pm \pi$, as it corresponds to   points of maximum curvature 
of $f(w)$.  Hence, there is an issue with the bias as well although---like 
in the variance---it does not affect the rate of convergence; it just affects the
constants in the bias and variance asymptotic expansions. 
All this goes to show that estimating $  { f}  (w)$ for $w =0$ or $\pm \pi$
may require more scrutiny, especially since estimating $  { f}  (0)$ is  
fundamental  for time series inference.  

To give a different analogy, in regression we often fit linear trends, globally
(as in straight line regression) or locally (as in local linear fitting of nonparametric
regression). If it is known that a regression function has a quadratic behavior
(either globally or locally), then it behooves us to fit a quadratic instead.
For example, in looking for the maximum (or minimum) of a regression function,
it is standard practice to fit a (local) quadratic; see e.g. Myers et al. (2016).  
Hence, our proposal of local quadratic fitting to estimate $  { f}  (w)$ for $w  $ 
at the boundaries is   natural, since we know that   $  { f}  (w)$ 
has a local maximum or minimum  for $w    =0$ or $\pm \pi$.  
 As we will see in Section \ref{sec.3}, the above simple Algorithm 
of local quadratic fitting    leads to a bias of   
order $O(M^{-4})$ at the boundary points, thus 
  taking advantage of  the aforementioned   boundary {\it opportunity.}

\end{Remark}

If Assumption A($p$) holds with $p>2$, then
a higher order polynomial (with only even powers) could also be used instead of the simple quadratics
(\ref{eq.loc-quad 0}) and (\ref{eq.loc-quad pi}), the goal again being to estimate the
respective intercept terms;  the details are straightforward and thus omitted.

\subsection{Improved spectral   estimation near the two boundaries}
\label{sec.2.3}

The dichotomous asymptotic behavior  of $\hat f(w) $ according to whether
$w=0$ or not obfuscates the need for a special treatment of $\hat f(w ) $
where $w$ is very close to zero (or $\pi$ for that matter).
For example, the estimator $\hat f(w_1) $ faces the same boundary problems
described in eq. (\ref{eq.one sided}), namely $\hat f(w_1) $ is (for the most part)
a one-sided average. Consequently,
$Var [\hat f(w_1)]\approx Var [\hat f(0)] $, which is double   the variance of
$\hat f(w) $ at a point $w$ far from the boundary.

Such boundary issues are shared by all $\hat f(w_j ) $ for $j=1, \ldots, m$, as well as all
  $\hat f(w_j)  $  with indices corresponding to the $m$ largest elements of set $J_n$;
here, $m$ is as defined in Algorithm \ref{algo.1}(i).
However, the fitted equations (\ref{eq.loc-quad 0}) and (\ref{eq.loc-quad pi}) give us a
way to address all boundary issues in a unified way.
We now define an estimated function $\tilde  f (w) $ for all $w\in [-\pi, \pi]$ by the  following
construction.
We first define it over the non-negative Fourier frequencies, so let:
\begin{equation}
\tilde  f (w_j)  = \begin{cases}
    \hat a_0+\hat b_0  w_j^2 \quad  \mbox{if}  \,  j = 0, 1, \ldots, m \\
    0 \qquad \qquad \mbox{if} \, j = m+1, \ldots , [n/2]-m \\
   \hat a_1+\hat b_1(w_j-\pi)^2   \quad  \mbox{if} \, j = [n/2]-m+1, \ldots, [n/2].
   \end{cases}
\label{eq.tilde fwj}
\end{equation}
We then extend to negative Fourier frequencies by symmetry, i.e.,
 letting $\tilde  f (w_j) = \tilde  f (-w_j) $
if $w_j<0$. Finally, we extend to   all of $  [-\pi, \pi]$ as follows:
if $w$ is not a Fourier frequency, let $\tilde  f (w)=\tilde  f (w_{j^*})$ where
$w_{j^*}$ is the Fourier frequency closest to $w$.

\section{Asymptotic performance}
\label{sec.3}

The spectral density $f( w)$ is assumed to be smooth, i.e., possessing a number of
derivatives, as postulated in  Assumption A$(p)$ of  Section \ref{sec.2}.
Therefore,  $f( w)$  admits a Taylor expansion of an appropriate order
 at any  frequency $\theta $  of interest.
At the boundary points, i.e., when
   $\theta = 0$ or $ \pm \pi$, the spectral density is an even function of $w -  \theta $, and hence
 the Taylor expansion only involves even order terms. Assuming   Assumption A$(4)$,
 we can write
 \begin{equation}
 \label{eq:spec.expand}
  f( w) = f(\theta) + \frac{1}{2 }  f^{(2)} (\theta) \, {(w -  \theta )}^2
   +   \frac{1}{4!} f^{(4)} (\theta) \, {(w -  \theta )}^4 + o ( {(w -  \theta )}^4 ),
\end{equation}
 where $o(1)$ denotes terms tending to zero as $ |w- \theta| \tends 0$.
 We will use this expansion to analyze the bias and variance of the estimator $\tilde f (w)$,
  focusing on the cases that $\theta = 0 $ or $ \pi$.


\subsection{Fitting via Ordinary Least Squares (OLS)}

 The Ordinary Least Squares (OLS)
 regression estimator  of $a_0$ and $b_0$  (or $a_1$ and $b_1$)
 at any of the boundary frequencies $\theta$ takes the form
 $ {( {\bf X}^{\prime} {\bf X} )}^{-1} {\bf X}^{\prime} {\bf Y}$,  where
 ${\bf X}$ has a first column of $m$ ones and a second column of $m$ values
  ${(  \theta  - w_j)}^2$.
    These $w_j$ are the Fourier frequencies
   falling in the interval $[  \theta  - 2 \pi \delta,  \theta  + 2 \pi \delta] \cap ( 0, \pi]$.
   Also ${\bf Y}$ is  a vector with entries $I(w_j)$, the periodogram evaluated at this
   band of Fourier frequencies.    To determine the Fourier frequencies that are used,
   note that by even symmetry at the boundaries we only need to focus on the interval
   $[0, \pi]$, and the scenario for $\theta = - \pi$ is the same as for $\theta = \pi$.
    Hence, for $\theta = 0$ we consider the set of $w_j \in (0, 2 \pi \delta]$,
    or $1 \leq j \leq m$ with $m =  [ \delta n]$.
    Note that  we don't use $j =0$ in the regression,  because typically the periodogram
    will have been mean-centered, implying that $I(w_0) = 0$.
  For $\theta =   \pi$,  we consider the set of $w_j \in [  \theta  - 2 \pi \delta, \pi]$,
  or $ [ n /2] - m +1  \leq j \leq [n/2]$.   We can take the lower bound to
  be  $[n/2 - n \delta]$  instead of  $[n/2] - m+1$  for purposes of asymptotic analysis.

 Now $w_j  =  2 \pi j/n$, and we will approximate $ {( {\bf X}^{\prime} {\bf X} )}^{-1}$
  using summation formulas;  for either $\theta = 0$ or $\theta = \pi$ we have
   the approximation (it is exact if $\theta = 0$)
\[
   {\bf X}^{\prime} {\bf X}  =  \left[ \begin{array}{cc}
  	m &  \sum_j  {(\theta - w_j) }^2 \\
  	 \sum_j  { (\theta - w_j) }^2 &  \sum_j  { (\theta - w_j) }^4  \end{array} \right]
  \approx   \left[ \begin{array}{cc}
  	m &  {(2 \pi/n)}^2  \sum_{j=1}^m  j^2 \\
  	 {(2 \pi/n)}^2  \sum_{j=1}^m  j^2  &   {(2 \pi/n)}^4  \sum_{j=1}^m  j^4   \end{array} \right].
\]
 Taking the limit as $n \tends \infty$ of $m^{-1} {\bf X}^{\prime} {\bf X}$ yields the matrix $Q$:
 \[
   Q =  \left[ \begin{array}{cc} 1 & {(  2 \pi\delta ) }^2  /3 \\
        {(  2 \pi\delta ) }^2  /3  &   {(  2 \pi \delta ) }^4  /5 \end{array} \right].
  \]
   Similarly,
\[
    m^{-1} {\bf X}^{\prime} {\bf Y} =
    \left[ \begin{array}{c}  m^{-1} \sum_j  I (w_j)  \\  m^{-1} \sum_j  I(w_j)  {(\theta -w_j)}^2 \end{array} \right].
\]
Let the region $A$ denote either $[0, 2 \pi \delta]$ or $[\pi - 2 \pi \delta, \pi]$, depending on whether $\theta = 0$
 or $\theta =   \pi$.   Then asymptotically the estimator takes the form
  \begin{eqnarray*}
&   \tilde f (w) & =  [ 1,  {( \theta  - w)}^2 ] \,   {( {\bf X}^{\prime} {\bf X} )}^{-1} {\bf X}^{\prime} {\bf Y}  \\
 &  & \approx   [ 1,  {( \theta  - w)}^2 ] \, Q^{-1}  \,  \left[ \begin{array}{c}  m^{-1} \sum_j  I (w_j)  \\
     m^{-1} \sum_j  I(w_j)  {( \theta  -w_j)}^2 \end{array} \right]  \\
 &  &=  m^{-1} \sum_j    [ 1,  {( \theta  - w)}^2 ] \, Q^{-1}  \,  { [ 1, {( \theta  -w_j)}^2  ]}^{\prime} \, I(w_j)  \\
  &    & \approx   \frac{1}{ 2 \pi  \delta} \int_{A}
       [ 1,  {( \theta  - w)}^2 ] \, Q^{-1}  \,  { [ 1,  {( \theta    - \lambda )}^2  ]}^{\prime} \, I(\lambda) \, d\lambda.
\end{eqnarray*}
  Comparing with Remark 9.8.2 of McElroy and Politis (2020),
it is apparent that  the above estimator is a
so-called {\it spectral mean}, i.e., has the form $\langle g_{2 \delta} ( \cdot, w)  I \rangle$
 with weighting function
\begin{equation}
\label{eq:ols-weight}
  g_{2 \delta} ( \lambda, w) =   \delta^{-1} [ 1,  {( \theta  - w)}^2 ] \, Q^{-1}  \,  { [ 1,  {(\theta  - \lambda )}^2  ]}^{\prime},
 \end{equation}
  which has been defined on $[0, \pi]$ only, i.e., on $[-\pi,0]$ it is defined to be zero.
Here,  $\langle \cdot \rangle$ denotes integration over $[- \pi, \pi]$, followed
by  division by $2 \pi$, so that
  $ \tilde f (w) \approx \langle g_{2 \delta} ( \cdot, w)  I \rangle$. 

  The asymptotic theory
  is then given by Theorem 9.6.6 of McElroy and Politis (2020), which assumes a linear process;
   however, cumulant conditions    as in  Brillinger (1981) can provide the same result.
 Let the tri-spectral density of the data process $\{ X_t \}$  be denoted by
\[
   F  (\omega, \lambda, \theta) = \sum_{h,k,\ell}  \gamma (h,k,\ell) \exp \{ -i (h \omega + k \lambda + \ell \phi) \},
\]
   where $\gamma (h,k,\ell) = \mbox{cum} ( X_t, X_{t+h}, X_{t+k}, X_{t+ \ell})$ is the fourth-order
   autocumulant function of the process.   The  following 
Central Limit Theorem (CLT) summarizes the asymptotic properties of 
       the estimator; the result is not limited
to $ g_{2 \delta} ( \lambda, w)$ having the form of eq. (\ref{eq:ols-weight}).

\begin{theorem}
\label{thm:local-spec-clt}
  Suppose $\{ X_t \}$ is a strictly stationary time series that is either a linear process  (with square summable
 MA($\infty$)  coefficients, and inputs with finite fourth moment)  or
   has autocumulant functions satisfying the summability condition (B1) of Taniguchi and Kakizawa (2000, p.55).
   Then, as $n \tends \infty$
\[
   \sqrt{n} \, (   \langle g_{2 \delta} ( \cdot, w)  I \rangle -  \langle g_{2 \delta} ( \cdot, w)  f \rangle )
   \convinlaw \mathcal{N} \left( 0, \langle  { g_{2 \delta}( \cdot, w) }^2 f^2 \rangle
    +  G   \right),
  \]
  \[ \mbox{where } \ \ 
 G=  \frac{1}{   {(2 \pi)}^{2} } \int \int g_{2 \delta}( \phi, w) g_{2 \delta}( \lambda, w)
      F (- \phi, \lambda, -\lambda) \, d\lambda d\phi .
\]
\end{theorem}

\paragraph{Proof of Theorem \ref{thm:local-spec-clt}.}
    Once the conditions are checked, this result directly follows from Lemma 3.1.1.(ii) of Taniguchi and Kakizawa (2000).
    Note that the first term in the asymptotic variance would normally also involve a term involving the integral
     of $g_{2 \delta} $ multiplied by its reflection about the y-axis; however, this product will be zero, since $g_{2 \delta}$
      is supported on $[0, \pi]$.   $\quad $ q.e.d.

  \begin{Remark} \rm
  \label{rem:local-spec-clt}
Focusing on  the case 
where $ g_{2 \delta} ( \lambda, w)$ has the form of eq. (\ref{eq:ols-weight}),
note that the centering in the above CLT is $\langle g_{2 \delta} ( \cdot, w)  f \rangle $ which 
is not equal to $f(w)$ in general; hence, there is an asymptotic bias term to be
  determined.  In addition, the asymptotic variance can be cumbersome since the quantity 
   $G$  depends on the tri-spectral density. Note that
   $G = 0$  if $\{ X_t \}$ is a  Gaussian process but it need not vanish in general. 
In particular, if $\{ X_t \}$ is a linear process with innovations of kurtosis $\eta$,
  we have $G = (\eta - 3)  { \langle g_{2 \delta}( \cdot, w) f \rangle }^2$; see
   Taniguchi and Kakizawa (2000, pp. 58-59). In this case,  $G$ can be estimated  by plugging in 
a consistent estimator of $f(w)$ in this expression.
 
\end{Remark}

   We now provide an asymptotic analysis of bias and variance in terms of a fixed $\delta $.
      The familiar situation where the bandwidth $\delta \tends 0$ will be studied  in the next subsection,
   where it is shown that $G$ is of smaller order (as $\delta \tends 0$) than the first term in the variance.  
  With this in mind, we  focus just  on the first term of the variance
   so as to develop an expression for the optimal bandwidth;
   a fixed bandwidth calculation can be most informative, as  
    the pioneering work of Kiefer   and  Vogelsang   (2002, 2005) has shown -- see also
   McElroy and Politis (2013, 2014). 
   For example, in our case at hand, 
    the  actual values of $\delta$ that minimize MSE can be as large as $0.15$,
    and therefore it might not be appropriate   
    to base the optimal selection of $\delta$ on formulas derived under a $\delta \tends 0$
    analysis. 

    Denote the entries of $m^{-1} {\bf X}^{\prime} {\bf X}$ by $c_0$ (upper left), $c_2$ (upper right), and
  $c_4$ (lower right).  For a non-negative integer  $p$ define
\[
   F_p = m^{-1} \sum_j {(\theta - w_j)}^p {f (w_j)}^2  
\ \ \mbox{and} \ \  G_p = m^{-1} \sum_j {(\theta - w_j)}^p f(w_j).
\]
%
%

\begin{proposition}
\label{prop:local-spec-biasandvar}
      Suppose $\{ X_t \}$  
 satisfies the conditions of Theorem \ref{thm:local-spec-clt}
      such that $G = 0$ in the asymptotic variance.   Assume (\ref{eq:spec.expand}),
and that $n \tends \infty$  but    $\delta $ is a fixed value. For
          $\theta = 0$ or $ \pi$ we have
\[
  \mbox{Var} [ \tilde{f} (\theta) ]  =   \delta^{-1}  n^{-1}
  \frac{ c_4^2 F_0 - 2 c_4 c_2 F_2 + c_2^2 F_4 }{ {(c_4 - c_2^2)}^2 }
  + o(n^{-1})
\ \ \mbox{and} \ \    \mbox{Bias} [ \tilde{f} (\theta) ]  = 
 \frac{ c_4 G_0 - c_2 G_2 }{ c_4 - c_2^2 }    -  f(\theta)  
  + O(n^{-1}).
  \]
\end{proposition}

\paragraph{Proof of Proposition \ref{prop:local-spec-biasandvar}.}
 Here we develop a variance  expression that has  a better finite-sample approximation then that given
  in Theorem \ref{thm:local-spec-clt}.
 Letting $[ a,  b] = [1, {(\theta - w) }^2 ] \,  { \left[  m^{-1} {\bf X}^{\prime} {\bf X} \right] }^{-1}$,
   we see the estimator is $\tilde{f} (w) = m^{-1} \sum_j (a + b {(\theta - w_j)}^2 ) I(w_j)$.
    Hence the variance is asymptotically given by
    $m^{-2} \sum_j { (a + b {(\theta - w_j)}^2 )}^2  {f(w_j) }^2$.
 Then
 \begin{eqnarray*}
  &  { \left[ m^{-1} {\bf X}^{\prime} {\bf X} \right] }^{-1}  & =    q^{-1} \,  \left[ \begin{array}{cc}
   c_4  & - c_2 \\   - c_2 & c_0   \end{array} \right]  \\
    & q & = c_4 c_0 - c_2^2 \\
    & [a,  b ] &  =  [ c_4 - c_2  {(\theta -w )}^2,  -c_2 + c_0 {( \theta - w)}^2 ]/q   \\
    &  a + b {(\theta - w_j)}^2  & =
    \left(  (c_4 - c_2 {( \theta - w_j )}^2 ) + {(\theta - w)}^2  (-c_2 + c_0 {(\theta - w_j)}^2 ) \right)/q.
\end{eqnarray*}
  Focusing on the case $w = \theta$, the asymptotic variance is
\[
 m^{-2} \sum_j {    (c_4 - c_2 {( \theta - w_j )}^2 ) }^2  {f(w_j) }^2 /q^2,
 \]
  which simplifies to the stated expression.
  To address the bias, let $\widetilde{\bf {Y}} $ be defined via replacing
  the periodogram by the true spectrum in ${\bf Y}$.
  So, taking the expectation of the estimator yields
  \begin{eqnarray*}
  &  E \widetilde{f} (w) & = [1, {(\theta - w)}^2] \, {( {\bf X}^{\prime} {\bf X} )}^{-1} {\bf X}^{\prime} E {\bf Y} \\
  &  & =  [1, {(\theta - w)}^2] \, {( {\bf X}^{\prime} {\bf X} )}^{-1} {\bf X}^{\prime} \widetilde{\bf Y}  + O(n^{-1}) \\
 & & = q^{-1} \,  [1, {(\theta - w)}^2] \, \left[ \begin{array}{cc} c_4 & - c_2 \\ -c_2 & c_0 \end{array} \right]
  \, \left[ \begin{array}{c} G_0 \\ G_2 \end{array} \right] +     O(n^{-1}).
\end{eqnarray*}
Setting $w = \theta$ yields the desired expression for the bias.  
$\quad$ q.e.d.

\vspace{.5cm}

  The Mean Squared Error (MSE) of $\tilde f(w)$ equals the squared bias plus the variance.
The value of $\delta$ that minimizes the   MSE is denoted $\delta_*$, and  can be calculated numerically given
the quantities $F_0$, $F_2$, $F_4$, $G_0$,  $G_2$, and $f(\theta)$.  However, these
 numbers are unknown in practice, and must be estimated.  A pilot estimator, such as the flat-top
 spectral estimator, can be used to estimate these quantities, in order that a numerically
  determined estimate of the optimal value of $\delta$ can be obtained.
  Such an empirical $\delta$ will be denoted by $\widehat{\delta}_*$; see Section \ref{sec.sims} for
   explicit details.


\subsection{Fitting via Weighted Least Squares (WLS)}
\label{se:WLS}

 The Weighted Least Squares (WLS) development generalizes the OLS approach.   We consider a kernel function $K$
 with domain $[0,1]$; usually the kernel is taken to be symmetric, but we focus on one-sided kernels due to our
 analysis at the boundaries of the frequency domain.   Define
   $K_{\delta} (x) = {(2 \pi \delta)}^{-1} K (|x|/ (2 \pi \delta))$, so that
  $K_{\delta}$ becomes a function on $[-\pi, \pi]$.  Then the weights are defined via
   $K_{\delta} ( \theta - w_j)$ for $1 \leq j \leq m$,
  and $ {\bf W}$ is a diagonal matrix consisting of these numbers.  The WLS estimator of $f(\theta)$ takes the
   form $ {( {\bf X}^{\prime} {\bf W} {\bf X} )}^{-1} {\bf X}^{\prime} {\bf W} {\bf Y}$.
   Setting   $K_{(j)} = \int_0^1 K (  u) u^j du$, and focusing on the cases that either $\theta = 0$ or $\theta = \pi$,
      we obtain
\[
 m^{-1}  {\bf X}^{\prime} {\bf W} {\bf X}   \approx
     \left[ \begin{array}{cc}   {(2 \pi \delta)}^{-1} K_{(0)} &   {(2 \pi \delta)} K_{(2)} \\
 	  {(2 \pi \delta)}  K_{(2)} &   {(2 \pi \delta)}^{3} K_{(4)}  \end{array} \right].
\]
  Let the limiting matrix be denoted $Q$.
   Similarly,
\[
    n^{-1} {\bf X}^{\prime} {\bf W}  {\bf Y} =
    \left[ \begin{array}{c}  m^{-1} \sum_j  K_{\delta} (\theta - w_j)  I (w_j)  \\  m^{-1} \sum_j  I(w_j)
     K_{\delta} (\theta - w_j)  {(\theta -w_j)}^2 \end{array} \right],
\]
 and  asymptotically the estimator takes the form of a spectral mean with weighting function
\begin{equation}
\label{eq:wls-weight}
  g_{2 \delta} ( \lambda, w) =  \delta^{-1}  [ 1,  {( \theta  - w)}^2 ] \, Q^{-1}  \,
    { [ 1,  {(\theta   - \lambda )}^2  ]}^{\prime}       K_{\delta} (\theta- \lambda).
 \end{equation}
  Then Theorem \ref{thm:local-spec-clt} holds for the WLS estimator, only replacing the weighting function
  (\ref{eq:ols-weight}) with (\ref{eq:wls-weight}).
   Evidently, setting $K(u) = 2 \pi \delta $ for $u \in [0,1]$ produces the OLS case.
   Likewise,  asymptotic bias and variance expressions akin to the OLS case can be worked out,
  which depend upon the function $K$.  In the following result we also include the role of $\delta$ as it tends to zero,
  corresponding to the classical case of spectral estimation. 

 \begin{proposition}
\label{prop:wls-spec-biasandvar}
      Suppose $\{ X_t \}$ 
satisfies the conditions of Theorem \ref{thm:local-spec-clt}.
        Let $K_{(j)} = \int_0^1 K (  u) u^j du$ and
    $\widetilde{K}_{(j)} = \int_0^1 { K(u)}^2 \, u^j \, du$.   Assume (\ref{eq:spec.expand}), and that $n \tends \infty$ with $\delta \tends 0$.   
  For $\theta = 0 $ or $ \pi$ we have
 \begin{eqnarray*}
 & \mbox{Var} [ \tilde{f} (\theta) ] & =  \delta^{-1} n^{-1}   {f( \theta)}^2 \, \left(
    \frac{  K^2_{(4)} \widetilde{K}_{(0)}  - 2   K_{(4)} K_{(2)} \widetilde{K}_{(2)} +
     K_{(2)}^2 \widetilde{K}_{(4)}  }{ {(K_{(0)} K_{(4)} - K_{(2)}^2 )}^2 }  + O(\delta^{-1} n^{-1}) \right) \\
&   & +  n^{-1}  F(-\theta, \theta, -\theta) \,   \frac{ \left( K_{(4)}^2 \widetilde{K}_{(0)}^2  -
 	 2 K_{(4)} K_{(2)} \widetilde{K}_{(2)} \widetilde{K}_{(0)} + K_{(2)}^2 \widetilde{K}_{(2)}^2 \right) }{
 	  {(K_{(0)} K_{(4)} - K_{(2)}^2 )}^2 }    + o(n^{-1})
  \end{eqnarray*}
\[
\mbox{and} \ \   \mbox{Bias} [ \tilde{f} (\theta) ]  =  -  \frac{   f^{(4)} (\theta) }{ 24}  {( 2 \pi \delta)}^4
    \,  \left( \frac{ K_{(4)}^2 - K_{(6)} K_{(2)} }{ K_{(4)} K_{(0)} - K_{(2)}^2 } \right)   + o(\delta^4) +  O(n^{-1}).
  \]
\end{proposition}

\paragraph{Proof of Proposition \ref{prop:wls-spec-biasandvar}.}
   First, note that $\det Q = {(2 \pi \delta)}^2  (K_{(0)} K_{(4)} - K_{(2)}^2 )$, and
\[
   g_{2 \delta} ( \lambda, w) =
     \frac{K_{\delta} (\theta - \lambda) }{ \delta \det Q} \,
       \left( {(2 \pi \delta)}^3  K_{(4)}   -   {(2 \pi \delta)}  K_{(2)} \,
        ( {( \theta  -w )}^2 + {( \theta  -\lambda )}^2 )
       +  {(2 \pi \delta)}^{-1}  K_{(0)} \,  {( \theta  -w )}^2 {( \theta  - \lambda )}^2 \right).
\]
     To get the first term in the  asymptotic variance, we square this expression and integrate over $A$ times ${f(\lambda)}^2$,
      dividing by $ 2 \pi$.      We focus on  the case that $w =  \theta $:
\[
  g_{2 \delta} (\lambda,  \theta ) =
      \frac{K_{\delta} (\theta - \lambda) }{ \delta  {(2 \pi \delta)}   (K_{(0)} K_{(4)} - K_{(2)}^2 ) } \,
     \left( {(2 \pi \delta)}^2  K_{(4)}   -    K_{(2)} \,          {( \theta  -\lambda )}^2    \right).
\]
     For $p = 0, 2,4$ we have as $\delta \tends 0$
\begin{eqnarray*}
   \int_A  {(  \theta  - \lambda )}^p  { f (\lambda )}^2 \,  { K_{\delta} (\theta - \lambda) }^2 \,
  d\lambda
   & = ( 2 \pi \delta )  \int_0^{ 1}  {( 2 \pi \delta u)}^p
     { f ( 2  \pi \delta u  + \theta  ) }^2 \, {K_{\delta} (- 2 \pi \delta u)}^2 \,    du \\
     & \approx    {(2 \pi \delta )}^{p-1}  {f (\theta)}^2  \, \widetilde{K}_{(p)},
\end{eqnarray*}
    using the change of variable $\lambda = 2 \pi \delta u +  \theta $.
    The   first term in the asymptotic variance is $n^{-1}$ times
\begin{eqnarray*}
& &   \frac{   \int_A \left( {(2 \pi \delta)}^4  K_{(4)}^2 - 2  {(2 \pi \delta)}^2  K_{(4)} K_{(2)} {(\theta - \lambda)}^2
   + K_{(2)}^2 {( \theta - \lambda )}^4 \right)   { f (\lambda )}^2 \,
   { K_{\delta} (\theta - \lambda) }^2  d\lambda }{ \delta {( 2 \pi \delta)}^3  {  (K_{(0)} K_{(4)} - K_{(2)}^2 )}^2 } \\
      &    & \approx   \delta^{-1}  {f( \theta)}^2  \, \frac{ \left(  K^2_{(4)} \widetilde{K}_{(0)}
  - 2   K_{(4)} K_{(2)} \widetilde{K}_{(2)} + K_{(2)}^2 \widetilde{K}_{(4)} \right)  }{ {(K_{(0)} K_{(4)} - K_{(2)}^2 )}^2 }.
\end{eqnarray*}
 For the second term in the asymptotic variance, we can similarly derive  (for $p,q \in \{0,2,4\}$) as $\delta \tends 0$
 \begin{eqnarray*}
  & & \int_A \int_A {(  \theta  - \lambda )}^p  \, {(\theta - \phi)}^q  \,  { K_{\delta} (\theta - \lambda) }
   \,  { K_{\delta} (\theta - \phi) } \,  F(-\phi, \lambda, -\lambda)   \,  d\lambda d\phi  \\
   & & = {( 2 \pi \delta )}^2  \int_0^{ 1} \int_0^1  {( 2 \pi \delta u)}^p \,  {(2 \pi \delta v)}^q
      \, {K_{\delta} (- 2 \pi \delta u)}   \, {K_{\delta} (- 2 \pi \delta v)}  \, 
      F(-\theta - 2 \pi \delta v, \theta + 2 \pi \delta u, -\theta - 2 \pi \delta u) \,    du  dv\\
     &  & \approx    {(2 \pi \delta )}^{p+q}  \, F(-\theta, \theta, -\theta)  \, \widetilde{K}_{(p)} \, \widetilde{K}_{(q)},    
     \end{eqnarray*}
using the change of variable $\lambda = 2 \pi \delta u +  \theta $ and   $\phi = 2 \pi \delta v +  \theta $. 
Therefore,  applying this yields
\begin{eqnarray*}
 G & = &    {(2 \pi \delta)}^{-4}   {  (K_{(0)} K_{(4)} - K_{(2)}^2 )}^{-2}
  \,  \int_A \int_A    { K_{\delta} (\theta - \lambda) }
   \,  { K_{\delta} (\theta - \phi) } \,  F(-\phi, \lambda, -\lambda)    \\ 
   & & \cdot \left( {(2 \pi \delta)}^4  K_{(4)}^2 - 
     {(2 \pi \delta)}^2  K_{(4)} K_{(2)} [ {(\theta - \lambda)}^2 + {(\theta - \phi)}^2 ] 
   + K_{(2)}^2 {( \theta - \lambda )}^2  {( \theta - \phi )}^2 \right)       \,  d\lambda d\phi   \\
   & \approx &  F(-\theta, \theta, -\theta) \,   \frac{ \left( K_{(4)}^2 \widetilde{K}_{(0)}^2  -
 	 2 K_{(4)} K_{(2)} \widetilde{K}_{(2)} \widetilde{K}_{(0)} + K_{(2)}^2 \widetilde{K}_{(2)}^2 \right) }{
 	  {(K_{(0)} K_{(4)} - K_{(2)}^2 )}^2 }.
\end{eqnarray*}   
  Now we turn to the bias; from Theorem \ref{thm:local-spec-clt},
  asymptotically the    expected value of the estimator is 
\[
\langle g_{2 \delta} ( \cdot, w)  f \rangle  =  
  \langle g_{2 \delta} (\cdot, w) \left(  f(\theta) + \frac{1}{2 }  f^{(2)} (\theta) \, {(\cdot -  \theta )}^2  \right) \rangle
  +   \langle g_{2 \delta} (\cdot, w) \,
   \left(  \frac{1}{4!} f^{(4)} (\theta) \, {(\cdot -  \theta )}^4  + o({(\cdot -  \theta )}^4) \right)
    \rangle,   
\]
via   (\ref{eq:spec.expand}).   By (\ref{eq:wls-weight}) the first term on the right hand side is
 \begin{eqnarray*}
& &  \delta^{-1}  [ 1,  {( \theta  - w)}^2 ] \, Q^{-1}  \,  \langle 
{ [ 1,  {(\theta   - \cdot )}^2  ]}^{\prime}   \, K_{\delta} (\theta - \cdot) \,   [1,  {(\cdot  -  \theta  )}^2 ] \rangle \,
   { [f (\theta),     f^{(2)} (\theta)/2 ]}^{\prime}  \\
&  &  =   [ 1,  {( \theta  - w)}^2 ] \, Q^{-1}  \, Q  \,
   { [f (\theta),     f^{(2)} (\theta)/2 ]}^{\prime}  \\
& & =    f(\theta) + \frac{1}{2 }  f^{(2)} (\theta) \, {(w -  \theta )}^2,
 \end{eqnarray*}
  using  -- from  the Taylor expansion  (\ref{eq:spec.expand}) --
\[
   f(\theta) + \frac{1}{2 }  f^{(2)} (\theta) \, {(\lambda -  \theta )}^2  =
    [1,  {(\lambda  -  \theta  )}^2 ] \, { [f (\theta),     f^{(2)} (\theta)/2 ]}^{\prime}.
\]
  Plugging in $w = \theta$, we find that the asymptotic bias is given, up to terms that are $o( \delta^4)$, by
\begin{eqnarray*}
& \approx &    - \frac{1}{ 4!}  f^{(4)} (\theta)  \, [1,0] Q^{-1}
  { \left[  m^{-1} \sum_j {(\theta - w_j)}^4 K_{\delta} (\theta - w_j),
   m^{-1} \sum_j {(\theta - w_j)}^6 K_{\delta} (\theta - w_j) \right] }^{\prime} \\
   & \approx &    - \frac{1}{ 4!}  f^{(4)} (\theta)  \,
   \frac{1}{ \det Q }  [ {(2 \pi \delta)}^3 K_{(4)},  - (2 \pi \delta) K_{(2)} ]  \,
     { [ {(2 \pi \delta)}^3 K_{(4)},  {(2 \pi \delta)}^5 K_{(6)} ] }^{\prime} \\
& = &      - \frac{1}{ 4!}  f^{(4)} (\theta)  \,  {( 2 \pi  \delta)}^4    \,
     \frac{ K_{(4)}^2 - K_{(6)} K_{(2)} }{ K_{(4)} K_{(0)} - K_{(2)}^2 },
\ \ \mbox{ as desired.}  \ \ \mbox{q.e.d.}
\end{eqnarray*}

\vskip .1in
\noindent   Proposition \ref{prop:wls-spec-biasandvar} implies the
consistency of the WLS estimator $\tilde{f} (\theta)$; it also implies the
consistency of the OLS estimator since the latter is a special case of the WLS,
with all weights being equal.

\section{Positivity and further issues}
\label{sec.log}

\subsection{Positive spectral   estimation near the two boundaries}
\label{sec.2.4}

As discussed in Remark \ref{Re.hat f plus}, the flat-top estimator $\hat f  (w)$
is not guaranteed to be non-negative---let alone positive; the same is true for
the new estimator $\tilde  f (w)$.
One can take the positive part estimator, i.e., define
$\tilde f ^+ (w)= \max (\tilde f  (w),0)$ for all $ w\in [-\pi, \pi]$,
by analogy to the modification on $\hat f  (w)$; this is tantamount to
using  OLS or WLS under the linear constraint $   f (w)\geq 0$  to
perform the local quadratic regression
that estimates $  f (w)$.

However, there is an important application where such a modification poses a problem.
It can be easily calculated that $Var [\sqrt{n}\bar X_n] \to f(0)$ as $n\to \infty$;
see Ch. 9 of McElroy and   Politis  (2020). Hence, in order to conduct inference for the
mean, $f(0)$ must be estimated in a strictly positive fashion, since estimating
$f(0)$ by zero has serious repercussions for  the underlying data, e.g. they may be
{\it over-differenced}; see McElroy and   Politis  (2013).

A simple resolution is to choose  a negligible (but nonzero) lower threshold for $\tilde f  (w)$,
as suggested by Politis (2011)  and also by
McMurry and  Politis (2015). To elaborate, let $\varepsilon$ be some positive number, and define
$\tilde f ^\varepsilon  (w)= \max (\tilde f  (w),  \varepsilon /n )$ for all $ w\in [-\pi, \pi]$. Whatever
the choice of $\varepsilon$ may be, the division by $n$ makes this modification asymptotically
negligible when $f  (w)>0$, i.e., $\tilde f ^\varepsilon  (w)$ and $\tilde f  (w)$ are
asymptotically equivalent.

To propose a different fix, let $\epsilon_j=I(w_j)/f (w_j)$ for $0<j<[n/2]$, and recall that
the $\epsilon_j$ are (approximately) i.i.d.~Exponential with mean one.
Then, $\log I(w_j)=\log f (w_j)+\log \epsilon_j$.
Since $E\log \epsilon_j\approx 0.57721$ (the Euler constant), we are led to the
homoscedastic regression
\begin{equation}
\log I(w_j)=\log f (w_j)+0.57721 +Z_j,
\label{eq.log I reg}
\end{equation}
 where the $Z_j= \log \epsilon_j-0.57721 $
are (approximately) i.i.d.~with mean zero; by contrast, the
regression of $I(w_j) $ on $ f (w_j)$ we employed in Section \ref{sec.2.2}
was heteroscedastic.

Wahba (1980) used regression (\ref{eq.log I reg}) in order to fit a spline to
the log-periodogram. We instead employ  the local quadratic approach focusing on the
two boundaries only. By analogy to (\ref{eq.loc-quad 0}) and (\ref{eq.loc-quad pi}), we
  write the Taylor approximations:
\begin{equation}
 \log f(w)\approx A_0+B_0w^2  \  \  \mbox{for} \  \ w\in [-  2  \pi \delta,  2  \pi \delta]
\label{eq.loc-quad 0 LOG}
\end{equation}
 \begin{equation}
 \mbox{and} \ \ \log f(w)\approx A_1+B_1(w-\pi) ^2   \  \  \mbox{for} \  \ w\in [\pi- 2  \pi \delta, \pi+ 2  \pi \delta] .
\label{eq.loc-quad pi LOG}
\end{equation}
 \vskip .1in

Our proposal  based on log-periodogram regression    is given in  the  following algorithm:
\begin{algorithm}
\label{algo.2}
\vskip .1in
\end{algorithm}

\begin{enumerate}
\item [(i)]
Let $\hat A_0$ and $\hat B_0$
be  the estimators of  $A_0$ and $B_0$  in eq. (\ref{eq.loc-quad 0 LOG}) via  a
(possibly weighted)   regression based on eq. (\ref{eq.log I reg}). The data to be used in this regression
are $I(w_1), \ldots, I(w_m) $, where $w_m$ is the largest Fourier frequency less or equal to $ 2  \pi \delta$.

\item [(ii)]
Similarly,  $\hat A_1$ and $\hat B_1$
are  the estimators of  $A_1$ and $B_1$  in eq. (\ref{eq.loc-quad pi LOG})
via   a
(possibly weighted)   regression based on eq. (\ref{eq.log I reg}).
The data to be used in this regression
are the $I(w_j)  $  with indices corresponding to the $m$ largest elements of set $J_n$.

\item [(iii)] Finally,  $\bar f(0)=e^{\hat A_0}$  is  the proposed positive  estimator of $  f(0)$,
and  $\bar f(\pi) =e^{\hat A_1}$ is  the proposed positive estimator of $  f(\pi)$.

\end{enumerate}
\noindent
By analogy to Section \ref{sec.2.3}, we can extend the new estimator $\bar  f (w)$ to a function on
all of $[-\pi, \pi]$. As before,
we first define it over the non-negative Fourier frequencies, so let:
\begin{equation}
\bar  f (w_j)  = \begin{cases}
    \exp \{\hat A_0+\hat A_0 w_j^2\} \quad \mbox{if} \, j = 0, 1, \ldots, m \\
    0 \qquad \qquad  \mbox{if} \, j = m+1, \ldots , [n/2]-m \\
   \exp \{\hat A_1+\hat B_1(w_j-\pi)^2\}   \quad \mbox{if} \,  j = [n/2]-m+1, \ldots, [n/2].
\end{cases}
\label{eq.bar fwj}
\end{equation}
We then extend to negative Fourier frequencies by symmetry, i.e.,
 letting $\bar  f (w_j) =\bar f (-w_j) $
if $w_j<0$. Finally, we extend to   all of $  [-\pi, \pi]$ as follows:
if $w$ is not a Fourier frequency, let $\bar f (w)=\bar   f (w_{j^*})$ where
$w_{j^*}$ is the Fourier frequency closest to $w$.

 The asymptotic analysis of estimator $\bar  f (w_j)  $ parallels the technical arguments
given in Section~\ref{sec.3}, and is thus omitted. We will  compare $\bar  f (w_j)  $ to
$\tilde f (w_j)  $ using finite-sample simulations in Section~\ref{sec.sims}.

\subsection{Improved spectral   estimation for boundary and non-boundary points}
\label{sec.2.5}
The estimator  $\tilde  f (w)$   from  Section \ref{sec.2.3} and
the estimator  $\bar  f (w)$ from  Section \ref{sec.2.4}
were meant to optimally address points in  the  neighborhoods of the
boundaries. Away from the boundaries we already have excellent estimators, e.g. the
flat-top estimator $\hat  f ^+(w)$ from Section \ref{sec.2.1}.
We can combine the two to obtain an estimated spectral density function that is
accurate over all of $  [-\pi, \pi]$.
To that end, define a weight function $\kappa (w) $ on $  [-\pi, \pi]$ as follows:
\begin{equation}
 \kappa (w)  = \begin{cases}
    w/( 2  \pi \delta)  \quad \mbox{if} \,  w\in [0,  2  \pi \delta]  \\
   1  \qquad \qquad \mbox{if} \,  w\in [ 2  \pi \delta,\pi- 2  \pi \delta]  \\
     - (w-\pi )/( 2  \pi \delta)  \quad \mbox{if} \, w\in [ \pi-  2  \pi \delta,\pi],
     \end{cases}
\label{eq.kappa}
\end{equation}
where $\delta \in (0,.5)$ is small; also define $\kappa (w) =\kappa (-w) $
for $w<0$.
Our proposed estimator of $f (w) $ on the whole of  $  [-\pi, \pi]$ is then
\begin{equation}
\check f(w) = \frac{ \kappa (w) \hat  f ^+(w) + \left(1-\kappa (w)\right)\tilde f^+     (w)}{C},
\label{eq.check}
\end{equation}
where the constant $C$ is chosen to ensure that $(2\pi)^{-1}\int_{-\pi}^\pi \check f(w) dw$
equals $\hat \gamma (0)$.
Recall that $\gamma (0)$ is the variance of   $X_t$, for  which our preferred
estimator  is $\hat \gamma (0)$. It is therefore important that the estimated variance of   $X_t$
as implied by our spectral density estimator agrees with $\hat \gamma (0)$; see
Appendix~A of McMurry and  Politis (2015) for further discussion.  In other words,
 we wish to ensure that $C$ satisfies
$$ C= \frac{  \int_{-\pi}^\pi  \{\kappa (w) \hat  f ^+(w) + \left(1-\kappa (w)\right)\tilde  f ^+ (w)\}dw }{2\pi \hat \gamma (0)}. $$
Alternatively, $\check f(w)$ can be defined using either $\tilde  f^\varepsilon (w)$ or
$\bar  f (w)$ instead of $\tilde  f ^+(w)$ in eq.  (\ref{eq.check}), yielding an
estimator that is guaranteed to be positive at the origin.

\subsection{Going beyond stationarity}
\label{sec.4.3}
Note that our framework presumes  stationarity of the     process
generating the data $X_1,\ldots, X_n$, so that the spectral density of
$\{X_t\}$ 
is (a) well-defined, and (b) admits  an  expansion of the type   (\ref{eq:spec.expand}).
 It is possible to envision extensions  of our methodology to
piecewise     stationary or  piecewise
  locally stationary processes---see Zhou (2013, 2014). In such a case, our method could be
   applied to each of the   short windows of the sample where the data can be
thought to be approximately stationary.

The universe of nonstationary data generating processes is vast. Without
imposing some  structural  assumptions,  consistent estimation may be
simply impossible. For example, stationarity implies that $EX_t=\mu$
that does not depend on $t$. If we relax this assumption to 
$EX_t=\mu_t$  depending on $t$, then some structure has to be
imposed on $\mu_t$ to enable consistent estimation. 

To give a simple example of such    structure,   assume that the
nonstationary observations
  $W_1,\ldots, W_n$  are generated from the additive model
\begin{equation}
  W_t= \mu_t + X_t,
\label{eq.trend_model}
\end{equation}
where $\{X_t\}$  is a mean zero, stationary process with 
spectral density $f(\lambda)$. The values $X_1,\ldots, X_n$
are not observed directly but our objective is  still to 
estimate  $f(\lambda)$. 
Comparing to eq.~(1.4.1) of Brockwell and Davis (1991), model 
(\ref{eq.trend_model}) combines the trend and a possible seasonal effect
in the term $\mu_t$, which will be assumed to be deterministic and
 have a parametric functional form,
i.e., the whole sequence $\{\mu_t, t\in {\bf Z}\}$  would be known if  the value of
a finite-dimensional parameter $\eta$ were  known.
Typically, $\eta$ would be estimable at   $\sqrt{n}$ rate, allowing for 
 its estimator $\hat \eta$  to be plugged-in without disturbing the
asymptotics of the slower-converging spectral estimates.
Examples include:  

\begin{itemize}

\item {\bf Seasonality:} $\mu_t$ is periodic with (known) period $d$, 
i.e., $\mu_t=\mu_{t+d}. $ Then,
$\eta = (\mu_1, \ldots, \mu_d)$ and can
be estimated as in Paradigm 3.5.3 of  McElroy and
Politis (2020).  

\item {\bf Piecewise constant trend:} In the case of a single change point, $\mu_t=\mu^{(1)}$ for $t\leq $ some $m$, while $
\mu_t=\mu^{(2)}$ for $t> m$ as in the real data example of Section
\ref{se:GLOTI};   $m$ should be such that $m/n \to c \in (0,1)$. 
Here, $\eta = (\mu^{(1)}, \mu^{(2)})$ but multiple change points are also possible.

\item {\bf Regression:}
 Assume $\mu_t= \sum_{k=1}^d \eta_k g_k(t)$ 
where the $g_k(t)$ are some given functions; e.g. $g_k(t)=t^{k-1}$.
The vector parameter $\eta = (\eta_1, \ldots, \eta_d)$
is typically estimated via OLS regression.  
\end{itemize}
 \noindent
To deal  with this general setup, we will adopt the 
well-studied econometric literature; see e.g.~Andrews (1991), Hansen (1992), and Politis (2011). 
Denote $\mu_t$ as $\mu_t(\eta)$, and 
let $X_t(\hat \eta)= W_t - \mu_t(\hat \eta)$.
Note that $X_t(  \eta)= W_t - \mu_t(  \eta)=X_t$.
We will adopt assumption V2 of Hansen (1992), restated below:
\vskip .1in
\noindent
{\bf Assumption  V.} Let ${\cal H}  $ be a neighborhood of  
  the true value of the parameter $\eta \in {\bf R}^d$,
  and let $|| \cdot ||$ denote Euclidean norm. 
Assume:  
 (i) $\sqrt{n}(\hat \eta- \eta)=O_P(1)$; 
  (ii) $ \sup_{t\geq 1} E \left( \sup_{\zeta \in {\cal H} }
|| X_t(\zeta) || ^2 \right) < \infty$; and 
(iii) $ \sup_{t\geq 1} E \left(  \sup_{\zeta \in {\cal H} }
|| \frac{ \partial  }{\partial \zeta  } X_t(\zeta)|| ^2 \right) < \infty$ 
where $\frac{ \partial  }{\partial \zeta  }$ denotes the gradient in
${\bf R}^d$.

\begin{corollary}
 Consider observed data
  $W_1,\ldots, W_n$    from   model (\ref{eq.trend_model}). 
Compute $\tilde f (\theta)$  via WLS fitting as in Section 
\ref{se:WLS} but using $\tilde I(\lambda)=n^{-1}|\sum_{t=1}^n \tilde X_t (\hat \eta) e^{it\lambda} |^2$ instead of $  I(\lambda)$.
If Assumption V and the conditions of 
Proposition \ref{prop:wls-spec-biasandvar} hold true, then the
conclusions  of 
Proposition \ref{prop:wls-spec-biasandvar} hold true verbatim.
\end{corollary}

\section{Finite-sample simulations}
\label{sec.sims}

 We evaluate the proposed procedures for estimating the spectral density at frequency $\theta = 0, \pi$
  by comparing their performance against the state-of-the-art of nonparametric spectral estimates,
using either    a Parzen or a     flat-top lag-window.   In particular,
   we compute the new estimator $\widetilde{f} (\theta)$  via OLS
using  the data-based optimal bandwidth $\widehat{\delta}_*$; the latter
  is determined by using the flat-top tapered spectral estimator to estimate the quantities in
  Proposition \ref{prop:local-spec-biasandvar}, and then optimizing the
  Mean Squared Error (MSE) with respect to $\delta$ (see below).
  We also consider a range of fixed $\delta$ values in our simulations, letting this quantity range from
   $.005$ to $.250$ in $50$ increments.

   As a first  benchmark procedure, we also compute $\widehat{f} (\theta)$
  using a trapezoidal flat-top taper, with $c = .5$ and bandwidth $M$  determined by the
   ``Empirical Rule" of Politis (2003).  In particular, let $\widehat{q}$ be the smallest
   positive integer such that $| \widehat{\rho}_{\widehat{q}+k} | <  1.96  \sqrt{\log_{10} n/n}$
   for $k \leq 1 + 3 \sqrt{\log_{10} n}$, and set $M = \widehat{q} / c = 2 \sqrt{q}$.

  As a second benchmark procedure, we use the well-known Parzen estimator, with optimal 
   plug-in   bandwidth selected according to the procedure    described in Politis (2003).
    In particular, in formula (14) of Politis (2003) the bias and variance
constants are estimated by utilizing
    tapered estimates of  derivatives of the spectral density, using the same flat-top taper
    and bandwidth $M$ discussed 
above.

\subsection{Data-based selection of the bandwidth-type parameter $\delta$.}
The reason that the plug-in bandwidth selection procedure of Politis (2003) works
for 2nd order kernels is that,  under Assumption A(4),  the 
flat-top estimates are characterized by a faster rate of convergence than the
estimate in question, namely the Parzen. We can borrow these ideas in order to
estimate $\widehat{\delta}_*$ needed in our local quadratic fitting;
here, however, we would need to assume Assumption A(6) for 
our bandwidth selection procedure to work.

The details  are as follows:   first,
      use the flat-top taper estimate $\widehat{f} $ to compute
      $\widehat{F}_0$, $\widehat{F}_2$, $\widehat{F}_4$, $\widehat{G}_0$, and $\widehat{G}_2$,
       where
\[
   \widehat{F}_p = m^{-1} \sum_j {(\theta - w_j)}^p {\widehat{f} (w_j)}^2  
\ \ \mbox{and} \ \ 
   \widehat{G}_p = m^{-1} \sum_j {(\theta - w_j)}^p \widehat{f} (w_j).
\]
  Also compute $c_2 = m^{-1} \sum_j  {(\theta - w_j) }^2$ and $c_4 = m^{-1} \sum_j  {(\theta - w_j) }^4$.
  Then, construct variance and bias estimates using Proposition \ref{prop:local-spec-biasandvar},
  so that our estimate of the MSE is
\[
  \widehat{\mbox{MSE}}_{\delta} =
      \delta^{-1}  n^{-1}   \frac{ c_4^2 \widehat{F}_0 - 2 c_4 c_2 \widehat{F}_2 + c_2^2 \widehat{F}_4 }{ {(c_4 - c_2^2)}^2 }
   +  { \left( \frac{ c_4 \widehat{G}_0 - c_2 \widehat{G}_2 }{ c_4 - c_2^2 } -  \widehat{f} (\theta)     \right) }^2.
\]
 The above function can   be numerically optimized over $\delta \in [0, 1/2]$, with  the global minimum being  denoted $\widehat{\delta}_*$.

\subsection{Basic simulation: a Gaussian ARMA model}  For our first simulation study, we consider the Gaussian ARMA(1,1) process $\{ X_t \}$  satisfying
\[
   X_{t} -\phi \, X_{t-1} =  Z_{t} +\vartheta  \ Z_{t-1},
\]
where $Z_t\sim$ i.i.d. $N(0,1)$.
 A range of parameters is considered: $\phi \in \{ -.9, -.5, 0, .5, .9 \}$ and $\vartheta \in \{ -.8, -.4, 0, .4, .8 \}$,
  resulting in 25  ARMA processes.    For each such process we consider samples of size $n = 50, 200, 800$,
   and assess the competing methods according to Bias, Standard Deviation (SD), and Root Mean Squared Error (RMSE)
 across $10^4$ Monte Carlo  replications.
 
   The results are summarized in the first fifty tables and figures
  of Appendix B (in the Supplementary  Material).  Specifically, these tables 
   compare the Parzen tapered estimate, the flat-top tapered estimate,
  and the new estimator   $\widetilde f(w)$
  (using  $\widehat{\delta}_*$),  according to Bias, SD, and RMSE  over the three sample sizes;
   both frequency $0$ and $\pi$ are considered for the 25 ARMA processes.  
       Overall, the new estimator is superior or roughly comparable (in terms of RMSE)
        to the flat-top estimator in the majority of
       the 25 cases. Its performance against the Parzen estimator is not quite as impressive, but it is still better        in about two thirds of the cases (focusing on $n=800$).  
         Furthermore, the improvement in performance  can sometimes be dramatic.
Unsurprisingly, the cases          where the flat-top estimator substantially beats the local quadratic estimator occur 
         when the spectral density is close to being flat, e.g. in the case of a white noise process.
         
   \subsection{An illustration and further comparisons}
       For illustration we focus on the particular case of  spectral estimation at $\theta = 0$ for
  an  ARMA process with $\phi = .9$ and $\vartheta = .4$.
  Here the  local quadratic spectral estimator at $n= 800$ is superior to the flat-top estimator,
  and comparable to the Parzen estimator  (see Table \ref{table-main:gaussARMA54dgptheta.0}).
  Results for the local quadratic with three fixed choices of $\delta \in \{ .05, .10, .25 \}$ are also
  displayed, which demonstrates the trade-off of Bias and Standard Deviation (SD).

\begin{table}[ht]
\centering
\begin{footnotesize}
\begin{tabular}{|l|ccc|ccc|ccc|}
  \hline
 \multicolumn{1}{|l|}{} & \multicolumn{3}{c|}{$n=50$}  
   & \multicolumn{3}{c|}{$n=200$}    & \multicolumn{3}{c|}{$n=800$}   \\  \hline \hline
 Method  & Bias & SD & RMSE & Bias & SD & RMSE & Bias & SD & RMSE \\ 
  \hline
Parzen taper & -147.167 & 45.659 & 154.087 & -75.840 & 79.586 & 109.935 & -30.992 & 71.468 & 77.899 \\ 
  Flat-top taper & -138.823 & 55.966 & 149.679 & -56.383 & 98.019 & 113.079 & -13.223 & 81.932 & 82.993 \\ 
Local ($\widehat{\delta}_*$) Quadratic  & -146.831 & 48.517 & 154.639 & -74.806 & 82.201 &
 111.144 & -30.357 & 70.214 & 76.496 \\ 
 \hline
   Local  ($\delta = .05$) Quadratic  & -114.756 & 76.528 & 137.933  &  
   	-72.176 &  60.268 & 94.03 &  -63.456 & 31.801 &  70.978    \\ 
   Local  ($\delta = .10$) Quadratic  & -137.778 & 45.997 & 145.253 & 
   	-116.417 & 31.622 & 120.636  & -110.462 & 16.9 & 111.748   \\ 
   Local  ($\delta = .25$) Quadratic  & -169.057 & 17.648 & 169.976 &  
   	-158.544 & 12.862 & 159.065 &  -155.777 & 6.927 & 155.931   \\ 
\hline
   Local ($\delta = .05$) Constant  &  -151.628 & 32.248 &  155.019  & 
   	-123.36 &  28.113  & 126.523 &  -116.76 & 15.329 & 117.761  \\ 
   Local  ($\delta = .10$) Constant  & -166.246 & 19.648 & 167.404 &  
   	-154.766 & 14.216 & 155.417  & -151.356 & 7.716 & 151.552   \\ 
   Local  ($\delta = .25$) Constant  &  -183.354 & 7.679 & 183.515 &  
   	-178.331 & 5.698 & 178.422  & -176.981 & 3.09 & 177.008   \\ 
\hline
   Local  ($\delta = .05$) Quartic  & -87.769 & 130.868 & 157.575  &
   	-43.368 & 89.517 & 99.469 &  -34.88 & 45.635 & 57.439 \\ 
   Local  ($\delta = .10$) Quartic  & -115.221 & 74.947 & 137.452  &  
   	-87.752 & 48.438 & 100.233 &  -80.569 & 25.51 & 84.511   \\ 
   Local  ($\delta = .25$) Quartic  & -156.072 & 28.131 & 158.587 &
   	-140.86 & 20.085 & 142.285   & -136.945 & 10.742 & 137.365  \\ 
   \hline
\end{tabular}
\end{footnotesize}
\caption{\baselineskip=10pt Bias, Standard Deviation, and Root MSE  for spectral density estimators at frequency $\theta = 0$,
 for a Gaussian ARMA(1,1) process with $\phi = .9$ and $\vartheta = .4$. 
  Sample size is $n= 50,  200,  800$.  
  Flat-top tapered   estimation and Parzen-taper estimation are considered, with optimal bandwidth choices.
 Local quadratic spectral estimation  is      considered 
 with estimated optimal window $\widehat{\delta}_*$.}
\label{table-main:gaussARMA54dgptheta.0}
\end{table}

 We noted at the end of Section 2.2 that a higher order polynomial could be used in the regression,
  and here we explore this possibility by including a quartic term.  We also examine the effects of 
  restricting to a constant in the regression (i.e., omitting the quadratic term), which allows us to directly
  see the benefit of the quadratic regression over simple local constant fitting. 
For conciseness,  we focus   on the case of the Gaussian ARMA process with $\phi = .9$ and $\vartheta = .4$
  for these comparisons.  Simulation results are included in the nine lower rows of Table \ref{table-main:gaussARMA54dgptheta.0}.
 
As we do not have a data-based procedure for selecting the value of $\delta$ for the
quartic, we provide results for the three local polynomial estimators for
three  values of $\delta$; we   then compare the three estimators at their 
quasi-`optimal' $\delta$, meaning the best  of the three bandwidths (for each sample
size). 
It is not surprising that the local constant is inferior to   the quadratic (and the
quartic). In fact, the thesis of our paper is that local quadratic gives
advantages over kernel smoothing; 
recall that local constant fitting  using OLS   is   
 tantamount to kernel smoothing using the Daniell (1946)  kernel. 

In comparing the   quadratic to the
quartic, it appears that the former is preferable for $n=50$ or 200.
However, the quartic appears to gain a small edge for $n=800.$
Note that the quadratic works under assumption A(4),  although  the quartic
requires assumption A(6), under which it  may give an
asymptotic improvement in the order of the bias. 
In our example assumption A(6) is satisfied, so the
empirical results are coherent with our expectations. 

Nevertheless, there are several reasons that point to 
recommending the  local quadratic as sufficient for practical work. 
One is Ockham's razor, i.e., the principle of parsimony and the
recommendation to choose the simplest model, other things being equal. 
To name two additional reasons:
(a) the effect of using a higher-order polynomial, e.g. 4th order,
will be noticeable for very large samples and only 
 if  the  higher-order derivative
in question, e.g. $f^{(4)}(\theta)$ for the quartic, happens  to be large
(in absolute value); and 
(b) one needs a working data-based procedure for selecting the
optimal  $\delta$; at this point, we have such a procedure for the 
quadratic but not for higher-order polynomials.

  \begin{figure}[htb!]
\centering
\begin{tabular}{c}
\epsfig{file = 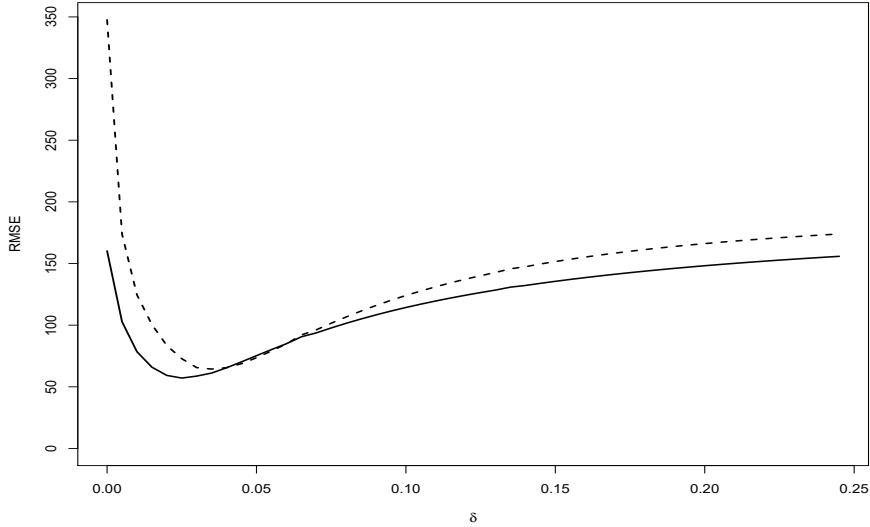,  angle = 0, height = 8cm, width = 12cm}
\end{tabular}
      \caption{\baselineskip=10pt  RMSE of Local Quadratic spectral estimation (solid) and Local
     Log Periodogram   estimation (dashed) at frequency $0$ plotted
        as a function of $\delta$, based on a sample of size $n=800$ of a
       Gaussian ARMA(1,1)        process    with $\phi = .9$ and $\vartheta = .4$.   }
     \label{fig-main:arma54.0}
 \end{figure}

\subsection{Comparison to local quadratic fitting of the  log-periodogram.}
It is also of interest to compare the     local quadratic   estimator $\tilde f(w)$
and the log-periodogram estimator $\bar f(w)$.
 The  first 50 figures of Appendix B  focus on the behavior of the new estimator as a function of $\delta$, letting this
  tuning parameter range over $(0, .25]$.  We compare the local quadratic estimator
$\tilde f(w)$  to the log-periodogram estimator $\bar f(w)$
  for sample size $n=800$, for all 25 ARMA processes at both frequency $0$ and $\pi$.
   An example, corresponding to  the ARMA process with $\phi = .9$ and $\vartheta = .4$, is seen
   in Figure  \ref{fig-main:arma54.0}.
 In general the log-periodogram quadratic estimator has inferior performance, although there are some values
  of $\delta$  for some of the processes in which the  estimator $\tilde f(w)$ had higher RMSE.
   However, such cases did not occur at an optimal $\delta$, i.e., when comparing the two curves at
    their respective minima,  the  log-periodogram quadratic estimator appears to be inferior in every case.
    This empirical result reinforces the intuitive idea that enforcing positivity at the outset entails a constraint that may hurt     performance. It is better to
correct for positivity after constructing an otherwise optimal estimator; this is the case
of  the local quadratic   estimator $\tilde f(w)$, as was the case of the
aforementioned flat-top tapered estimators.

\subsection{Simulation with non-Gaussian processes}
Finally, it is important to evaluate the new methodology  on non-Gaussian processes.  We consider three additional
 simulation exercises: (i) a non-Gaussian ARMA process driven by Laplace noise, (ii) a heavy-tailed ARMA process
   driven by Student $t$ noise with $6$ degrees of freedom, and (iii) a polynomial Gaussian process of order two. 
    Both the ARMA processes 
  have the same parameter ranges as used in the Gaussian case, and the inputs are normalized so as to have variance
  one.  The polynomial Gaussian process is a nonlinear process defined in Terdik and Meaux (1991), and is   
  constructed by passing Gaussian white noise through a quadratic filter -- see (A.1) of Appendix A.
  (We derive the polyspectra of this process in Proposition A.1, confirming its nonlinear properties; we also show
  that with a particular choice of the filter coefficients the spectral density has the form corresponding to
 the sum of two independent AR(1) processes, even though the process is nonlinear.)
 The two parameters, $\phi_1$ and $\phi_2$, that govern the dynamics of the polynomial Gaussian process
  are each allowed to range  in the set  $\{  -.9, -.5, 0, .5, .9 \}$.
  The results of these simulations are provided in the tables and figures of Appendix B; since the performance of
 the estimators is qualitatively similar to the Gaussian case, we do not present these results in the main paper.
 In general, the local quadratic estimator is superior except in the case that the true process is a white noise
 (i.e., has a flat spectral density at the boundary).

\newpage

\section{Real Data Examples}
 \label{se:gross}

We next study two data sets of current interest, U.S. Gross Domestic Product (GDP) and
  the Global Land-Ocean Temperature Index (GLOTI).

\subsection{U.S. Gross Domestic Product}

 \begin{figure}[htb!]
\centering
\begin{tabular}{c}
\epsfig{file = 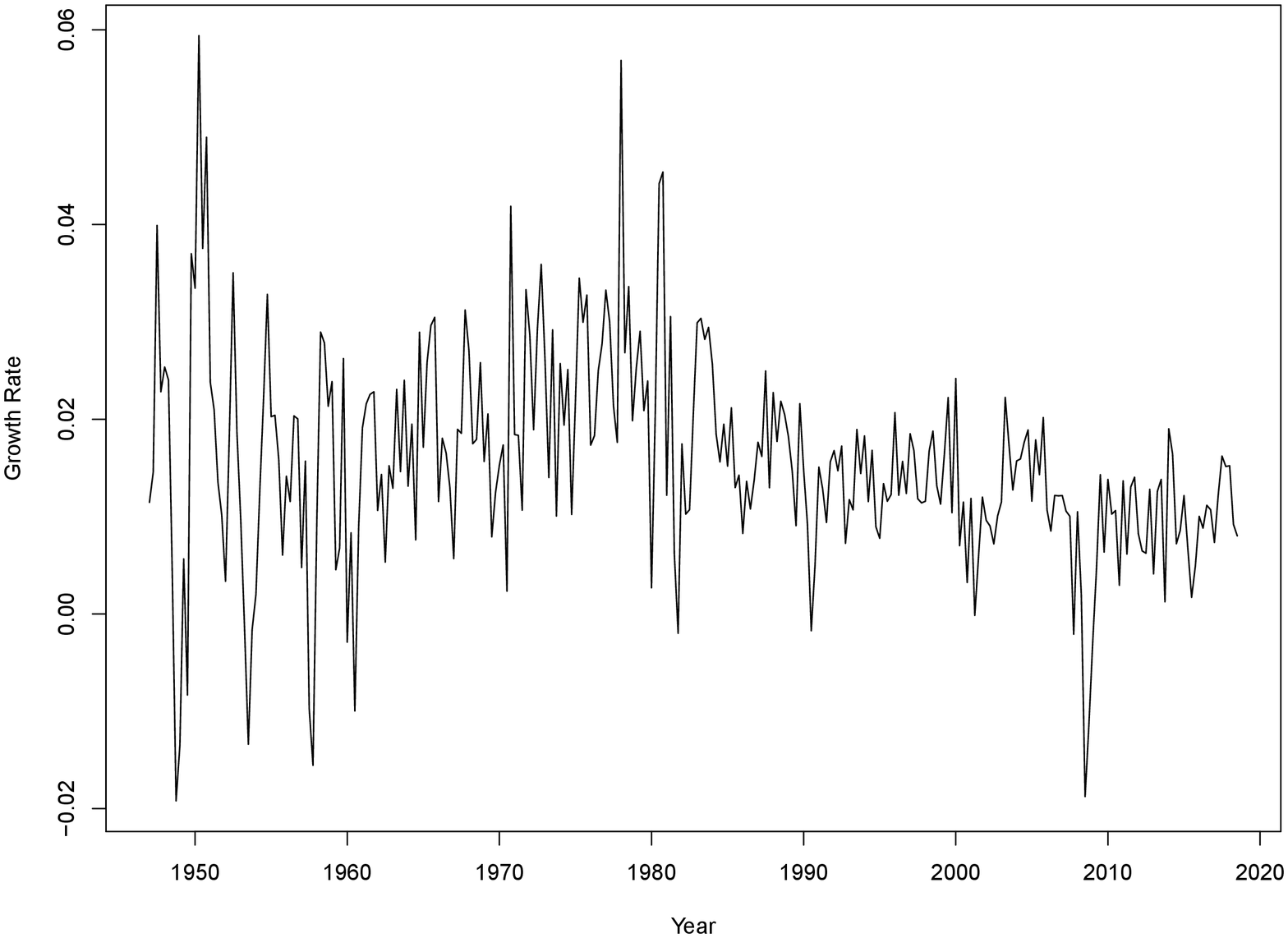,  angle = 0, height = 8cm, width = 12cm}
\end{tabular}
      \caption{\baselineskip=10pt  Growth rate of  U.S. Gross Domestic Product (GDP),
 from 1947Q1 through 2020Q3 (units of billions
 of dollars, seasonally adjusted annual rate).  Source: U.S. Bureau of Economic Analysis, Gross Domestic Product [GDP],
  retrieved from FRED, Federal Reserve Bank of St. Louis.    }
     \label{fig:gdp_growth}
 \end{figure}

We consider the time series  $\{ Y_t \}$  of quarterly GDP
 from 1947Q1 through 2020Q3 (units of billions
 of dollars, seasonally adjusted annual rate)\footnote{U.S. Bureau of Economic Analysis, Gross Domestic Product [GDP],
  retrieved from FRED, Federal Reserve Bank of St. Louis;
  \url{https://urldefense.com/v3/__https://fred.stlouisfed.org/series/GDP__;!!Mih3wA!VcvF56vPVXyxZNQZTiLhlZAYfL6m6vIdVZ4DS7pJ1Q60-k3FOtf9LQmDql_gcN4Zz1Q$ }, November 13, 2020.}.
  As one of the primary variables used to gauge the strength of the U.S. economy, GDP commands a tremendous interest;
   in particular, it is important to measure the  GDP growth rate.   Here we study the log differences of GDP $\{ X_t \}$,
    defined via $X_t = \log Y_t - \log Y_{t-1}$ for $ t = 1, 2, \ldots, n$, referred to as the GDP growth.   This GDP growth time series
    exhibits features that indicate stationarity (when restricting to sub-spans of the series)  is a reasonable hypothesis.
    In examining Figure \ref{fig:gdp_growth}  we see that the growth is sometimes negative, although in most quarters is positive;
    also there are possibly some cyclical effects present.

     It is natural to ask whether the mean growth rate is  within the typical range of growth, viz. $2 \%$ to $3 \%$
     (see Jones (2016) for background).  The answer
     clearly depends upon what time period is examined;  GDP growth rates
      are somewhat lower on  average  after 1980, and there seems to be a further moderation beginning around 2000.
      We can examine the hypothesis that GDP growth over the last 20 years exceeds the lower range of the typical
      growth rate, viz. $2 \%$; this corresponds to a one-sided
test of the null hypothesis   that GDP growth equals $.02$. Secondly, we can examine whether the growth exceeds the higher range of $3 \%$, i.e., a one-sided
test of the null hypothesis that GDP growth equals $.03$.

    In order to   test these hypotheses, we utilize the  mean growth rate of GDP  as a test statistic, restricting
     to the last 20 years  (or $n=80$) of
      quarterly observations.    As discussed in Section 4.1, the asymptotic
      variance of the sample mean of a stationary
      time series is $f(0)$ divided by sample size.
 We estimate $f(0)$ using the methods proposed in this paper,
       forming the $t$--statistic       $\sqrt{n} \overline{x}/ \sqrt{ f(0) }$, and using the standard normal asymptotic critical values.
       More specifically,  we consider both the local quadratic estimator and the
log-periodogram variant, and compare
       the results to using the flat-top and Parzen tapers (with optimal bandwidth selection, as described in the
        simulation studies).   Furthermore, we compare to estimates of $f(0)$ based on a fitted AR($p$) model,          where $p$ is determined by minimizing the AIC criterion, and the AR-fitting method is  via OLS.

The flat-top bandwidth is estimated to be $M=4$, resulting in a Parzen bandwidth of $31.82$.  The data-based selection of $\delta$  
 for the local quadratic estimator yields  $\widehat{\delta}_* =.245$.   The AR-model  method results in $\widehat{p}=2$, selected by AIC.
 The resulting estimates    are given in Table \ref{tab:gdp-results}.
  The largest estimate of $f(0)$ is given by the AR spectral estimator, followed by
the log-periodogram  estimator.  The
Parzen and local quadratic estimator give similar estimates, with 
the flat-top being also close. 

    \begin{table}[ht]
\centering
\begin{tabular}{|l|ccccc|}
  \hline
 &   Parzen  &   Flat-top   &     Local quadratic &   Log-periodogram  &      AR OLS  \\  \hline 
 \hline
spectrum      &  0.00011185 &  0.00012799 &  0.00011534  &   0.00013137  &  0.0001815  \\
$t$--statistic (null value $.02$)  & 4.34337242 &  4.06025137 &  4.27719210 &  4.00764423 &   3.4095815  \\
$t$--statistic (null value $.03$)   & 2.22905000 &  2.08375024 &  2.19508579  &  2.05675188  &  1.7498218 \\
Lower Limit &  0.03127268 & 0.03062629 &  0.03112925 &  0.03049612  &  0.02873390 \\
Upper Limit & 0.04981256 & 0.05045895 & 0.04995599 &  0.05058912 & 0.05235134 \\
 \hline
\end{tabular}
\caption{Results of analysis of GDP growth rate.}
\label{tab:gdp-results}
\end{table}

  In order to test the two hypotheses, we need to divide   each growth 
rate\footnote{The definition of annual
   growth rates by the U.S. Bureau of Economic  Analysis is given in
   \url{https://urldefense.com/v3/__https://www.bea.gov/help/faq/463__;!!Mih3wA!VcvF56vPVXyxZNQZTiLhlZAYfL6m6vIdVZ4DS7pJ1Q60-k3FOtf9LQmDql_gAGofiBQ$ }.}
by 4.  In the first case of $2 \%$ annual growth, an upper one-sided
   test leads to rejection for all of the estimates of $f(0)$.  In the second case of $3 \%$ annual growth,
   the story is different:  because of the higher estimate of $f(0)$ arising from the AR spectral estimator,
   the p-value is $.040$, indicating moderate evidence.  In contrast, the p-value for the local quadratic
   estimator's $t$--statistic    is $  0.014$, and the evidence is much more conclusive; this occurs because
   the relevant estimate of $f(0)$ turns out to be smaller---here is where
the improved    accuracy of local quadratic fitting can make  a difference in practice.

 We can also examine the annual growth over this period by constructing a $95 \%$ 
 confidence interval based on the asymptotic normality. 
Table \ref{tab:gdp-results} lists the lower and upper confidence limits
constructed via the five different methods.
 Whereas the AR method
   allows for a growth rate as high as $5.2 \%$,  the local quadratic estimator indicates the
    growth rate is strictly less than $5 \%$, but well above $3 \%$. 

\subsection{ Global Land-Ocean Temperature Index  }
\label{se:GLOTI}

 \begin{figure}[htb!]
\centering
\begin{tabular}{c}
\epsfig{file = 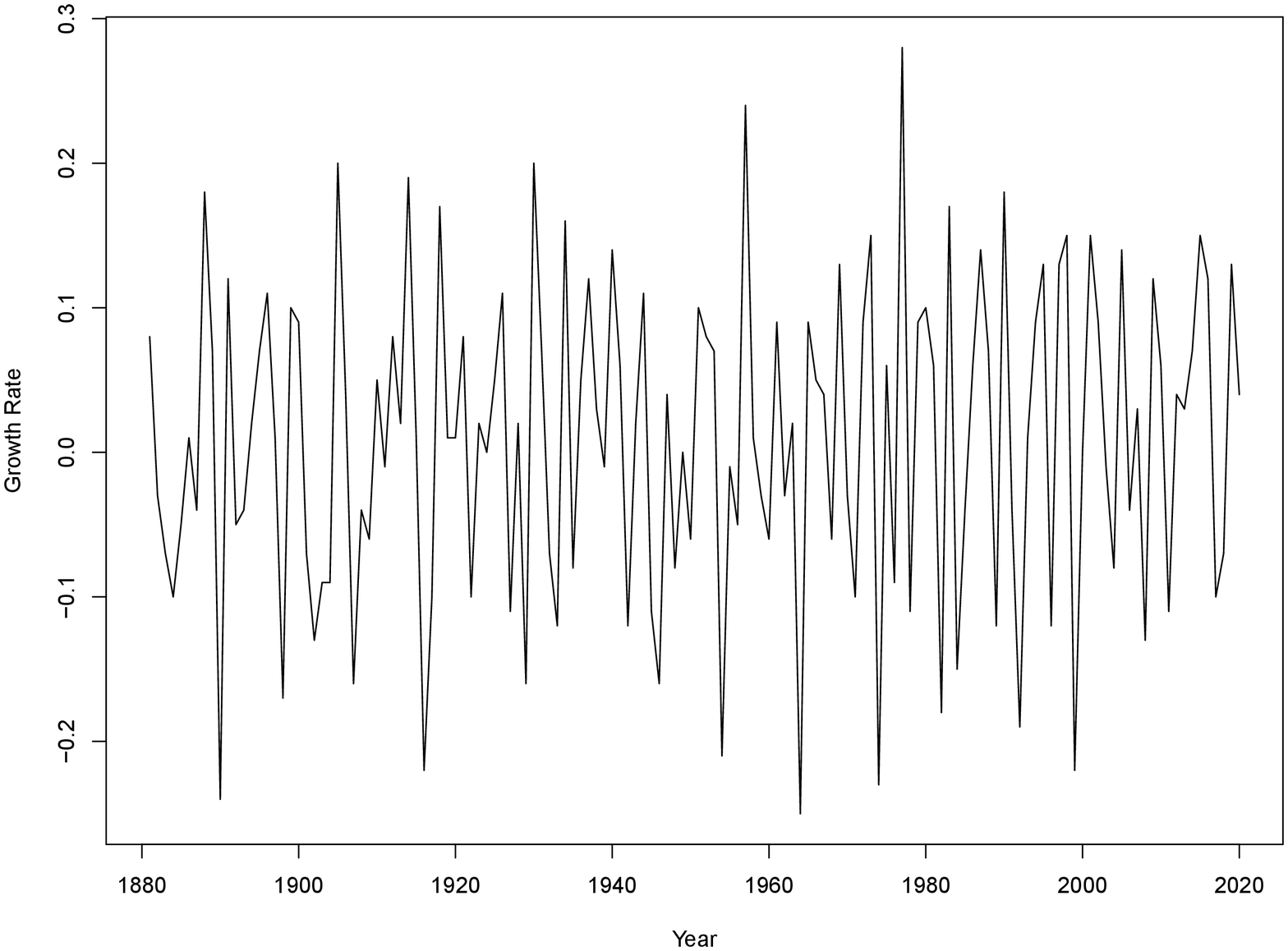,  angle = 0, height = 8cm, width = 12cm}
\end{tabular}
      \caption{\baselineskip=10pt  Growth rate of  Global Land-Ocean Temperature Index (GLOTI),
 from 1880 through 2020.  Source: NASA's Goddard Institute for Space Studies.    }
     \label{fig:temp_growth}
 \end{figure}
 
We study the time series $\{ Y_t \}$ corresponding to the GLOTI,\footnote{Data source: NASA's Goddard Institute for Space Studies, retrieved
from \url{https://urldefense.proofpoint.com/v2/url?u=https-3A__climate.nasa.gov_vital-2Dsigns_global-2Dtemperature_&d=DwIGAg&c=-35OiAkTchMrZOngvJPOeA&r=ArlQHWCRCNUBzA5G5_1B9GdyataLP0jyqZ2bB0rc--I&m=nJu0dJ8ItbpQqfaMV57vykhCYpgsE4ZFw-lCzJZjvzXh5KhdF30_iojNCT47iDed&s=fctvW9MgNu0kDnpzI2DITEgYI3hZfLJf30WboaljAAk&e= } on December 28, 2021.}  
an annual series covering the years 1880 through 2020.
 Climate data has received substantial attention over the past decades, and the GLOTI (in levels)
 shows a visually marked increase during the years following World War II.   The upward trend
 can be assessed by examining the mean of the differenced GLOTI series, defined as $X_t = Y_t - Y_{t-1}$
  for $t = 1,2, \ldots, n$.  This GLOTI growth series appears to have regular variation
  -- see Figure   \ref{fig:temp_growth} --   but it  is of interest to determine if the mean level   changed after World War II.  
    To address this question, 
consider a change point $t_*$ such that  
\begin{equation}
  E X_t =  \mu_1 1_{ \{ t  \leq t_* \} } +
       \mu_2 1_{ \{ t > t_* \} },
  \label{eq.two means}
\end{equation}
   and that $\{ X_t - E X_t \}$ is stationary; for simplicity,
we will further assume that  $\{ X_t - E X_t \}$ is $m$-dependent (for some $m > 0$).
 
  Our null hypothesis is $H_0: \mu_1 = \mu_2$, which is tested by comparing sample means
   $\overline{X}_1$ and $\overline{X}_2$ computed over samples before and after time $t_*$, respectively.
   That is, $\overline{X}_1$ is based on a sample of size $n_1$ with all times less than $t_* - m/2$, and
   $\overline{X}_2$ is based on a sample of size $n_2$ with all times greater than $t_* + m/2$.
   Then the $m$-dependence assumption ensures that $\overline{X}_1$ and $\overline{X}_2$ are independent.
   We can test $H_0$ with the statistic $\overline{X}_1 - \overline{X}_2$, which has asymptotic variance
   (under the null) of $(n_1^{-1} + n_2^{-1}) f(0)$, where $f$ is the spectral density of $\{ X_t - E X_t \}$.
   Hence, our studentized statistic is
\[
    \frac{ \overline{X}_1 - \overline{X}_2 }{ \sqrt{ (n_1^{-1} + n_2^{-1}) \hat{f}(0) }}.
\]

 Estimates $\hat{f} (0)$ of $f(0)$ can be obtained   using the entire sample
provided the data are centered correctly; for this reason, 
    we center the data by a mean estimate that 
remains consistent under $H_a: \mu_1 < \mu_2$
  (which states that growth rate has increased).  This is done by 
  estimating $\mu_1$ and $\mu_2$ via $\overline{X}_1$ and $\overline{X}_2$
respectively,  and plugging these estimates in eq. (\ref{eq.two means})
to get an estimate of $E X_t $ by which to center the data for 
  computing sample autocovariances, etc.
   As in our study of GDP,
   we consider the local quadratic estimator and the log-periodogram variant, as well as the flat-top,
    Parzen, and AR($p$) estimators.  In order to select the two samples, we set $t_*$ corresponding to
     the year 1950, demarcating the beginning  of the post-war industrial economy after recovery from the war years.
  Examination of the autocorrelation plot (not shown) indicates that   there is little substantial correlation past lag 2,
   and therefore we set $m=2$. 
   So with  $t_* = 70$  the first sample consists of $1 \leq t \leq 69$, and the second sample
   consists of $71 \leq t \leq 140$.

    \begin{table}[ht]
\centering
\begin{tabular}{|l|ccccc|}
  \hline
 &   Parzen  &   Flat-top   &     Local  quadratic &   Log-periodogram &      AR OLS  \\  \hline \hline
spectrum      &   0.00148086  &   0.00043289 &  0.00220545 &     0.00337354  &   0.00186396 \\
$t$--statistic  &   -2.51529769 & -4.65221263 & -2.06109805 &     -1.66649447 &  -2.24196225  \\
 \hline
\end{tabular}
\caption{Results of analysis of GLOTI growth rate.}
\label{tab:gloti-results}
\end{table}

The estimated flat-top bandwidth is $M = 2m = 4$, resulting in a Parzen bandwidth of 
$ 47.435$. 
  The optimal $\widehat{\delta}_*$
 for the local quadratic estimator is $0.02392 $.  
 The AR method results in $\widehat{p}=5$, selected by AIC.
 The resulting estimates    are given in Table \ref{tab:gloti-results}.
  The smallest estimate of $f(0)$ is given by the flat-top spectral estimator,  
   leading to a rejection of $H_0$ with p-value $1.64 \cdot 10^{-6}$.
  The Parzen,  AR , and local quadratic estimators yield larger
  p-values of $.00595$,  $.01248$,   and   $.01965$ respectively;
they are all  significant at level $0.05$.
  The log-periodogram variant is barely significant 
with p-value   $.04781$.  
 Overall there is support for the rejection of $H_0$ by all methods considered.

\section*{Acknowledgments}

This report is released to inform interested parties of research and to encourage discussion.  The views expressed on
statistical issues are those of the authors and not  those of the U.S. Census Bureau.  Research of the second
 author partially supported by NSF grant DMS 19-14556.
 

\end{document}